\documentclass{article} 
\usepackage{epsfig} 
\usepackage{cite}  
\def\sab{s_{12}}
\def\sac{s_{13}}
\def\sbc{s_{23}}

\def\H{\hbox{H}}

\def\d{\hbox{d}} 
 
\def\f{\hbox{f}} 
\def\FORM{{\tt FORM}} 
\def\g{\hbox{g}} 
\def\S{\hbox{S}} 
\def\Li{\hbox{Li}} 
\def\ln{\hbox{ln}}

\def\G{\hbox{G}}

\def\ie{i.e.} 
\def\iep{i\epsilon}

\begin{document} 
\unitlength1cm 
\begin{titlepage} 
\vspace*{-1cm} 
\begin{flushright} 
CERN-TH/2002-145\\ 
hep-ph/0207020\\
July 2002 
\end{flushright} 
\vskip 3.5cm 

\begin{center} 
{\Large\bf Analytic Continuation of\\[2mm] Massless 
Two-Loop Four-Point Functions}
\vskip 1.cm 
{\large  T.~Gehrmann}$^a$ and {\large E.~Remiddi}$^b$ 
\vskip .7cm 
{\it $^a$ Theory Division, CERN, CH-1211 Geneva 23, Switzerland} 
\vskip .4cm 
{\it $^b$ Dipartimento di Fisica, 
    Universit\`{a} di Bologna and INFN, Sezione di 
    Bologna,  I-40126 Bologna, Italy} 
\end{center} 
\vskip 2.6cm 

\begin{abstract} 
We describe the analytic continuation of two-loop four-point functions 
with one off-shell external leg and internal massless propagators
from the Euclidean region of space-like $1\to 3$ decay
to Minkowskian regions 
relevant to all $1\to 3$ and $2\to 2$ reactions with one space-like or 
time-like off-shell external leg. Our results can be used to 
derive two-loop master integrals and unrenormalized matrix elements 
for hadronic vector-boson-plus-jet production and deep inelastic 
two-plus-one-jet production, from results previously obtained  
for three-jet production in electron--positron annihilation. 
\end{abstract} 
\vfill 
\end{titlepage} 
\newpage 

\renewcommand{\theequation}{\mbox{\arabic{section}.\arabic{equation}}} 

\section{Introduction}
\setcounter{equation}{0}
In recent years, considerable progress has been made towards the 
extension of QCD calculations of jet observables towards the 
next-to-next-to-leading order (NNLO) in perturbation theory. 
One of the main ingredients in such calculations are the two-loop 
virtual corrections to the multi leg matrix elements relevant to 
jet physics, which describe either $1\to 3$ decay or $2\to 2$ scattering 
reactions: two-loop four-point functions with massless internal 
propagators and up to one off-shell external leg. 

Using dimensional regularization~\cite{dreg,hv} 
with $d\neq 4$ dimensions as regulator for
ultraviolet and infrared divergences, 
the large number of different integrals appearing in the 
two-loop Feynman amplitudes for $2\to 2$ scattering or $1\to 3$ decay 
processes 
can be reduced to a small number of master integrals. 
The techniques used in these reductions are 
integration-by-parts identities~\cite{hv,chet} 
and Lorentz invariance~\cite{gr}. A computer algorithm for the 
automatic reduction of all two-loop four-point integrals 
was described in~\cite{gr}. 

The use of these techniques 
allowed the calculation of  two-loop QED and QCD corrections to many $2\to 2$
scattering processes with massless on-shell external 
particles~\cite{m}, which require master integrals
corresponding to  massless four-point functions with all legs 
on-shell~\cite{onshell1,onshell2}. 
The results in~\cite{m} are given for all 
three physical Mandelstam channels, which are related by analytic 
continuation.
In~\cite{doublebox}, the full set of two-loop
four-point master integrals with one external leg off-shell 
was computed, for the kinematical situation of a $1\to 3$ decay,
by solving the differential equations in external invariants~\cite{gr} 
fulfilled by these master integrals. These 
integrals were employed in the calculation of the two-loop QCD corrections 
to the $e^+e^-\to 3$~jets matrix element and to the 
corresponding helicity amplitudes in~\cite{3jme}, which can be 
expressed as a linear combination (with rational coefficients in the 
invariants and the space-time dimension $d$) 
of the corresponding master integrals. 
The $2\to 2$ scattering
processes related to $e^+e^-\to 3$~jets by analytic continuation and 
crossing are both of high phenomenological importance:
hadronic vector-boson-plus-jet production and deep inelastic 
two-plus-one-jet production.

The two-loop four-point functions with all legs on-shell can be expressed 
in terms of Nielsen's polylogarithms~\cite{nielsen}, which have a 
well-defined analytic continuation~\cite{nielsen,bit}. The continuation of 
the master integrals in the on-shell 
case~\cite{onshell1,onshell2}
was discussed in detail 
in~\cite{onshell2}. In contrast, the closed analytic 
expressions for 
two-loop four-point functions~\cite{doublebox} with one leg off-shell contain 
two new classes of functions: harmonic polylogarithms (HPLs)~\cite{hpl,1dhpl}
and two-dimensional harmonic polylogarithms (2dHPLs)~\cite{2dhpl}. 
Accurate numerical implementations for HPLs~\cite{1dhpl} and 
2dHPLs~\cite{2dhpl} are available. 
The implementations apply 
to HPLs of arbitrary real arguments~\cite{hpl,1dhpl}, but only for a 
limited range of arguments of the 2dHPLs. This 
range of arguments corresponds precisely to the Euclidean region 
(space-like $1\to 3$ decay) for 
the corresponding off-shell master integrals~\cite{doublebox}. 
Continuation from the Euclidean region to any physical (Minkowskian) 
region requires in general the analytic continuation of the 2dHPLs 
outside the range of arguments considered in ~\cite{2dhpl}. Only 
in the special case of a time-like $1\to 3$ decay (relevant to 
$e^+e^-\to 3$~jets), which corresponds to the simultaneous continuation 
of all three external invariants from Euclidean to Minkowskian values, 
this continuation can be carried out by simply replacing an overall scaling 
factor, while preserving all 2dHPLs (which depend only on dimensionless 
ratios of the invariants). These results were used 
in~\cite{3jme} for the calculation of the two-loop matrix elements for  
$e^+e^-\to 3$~jets. 

It is the aim of the present paper to derive the relations 
for HPLs and 2dHPLs needed for the analytic continuation of the 
two-loop four-point master integrals of~\cite{doublebox} 
to the kinematics of all the $2\to 2$ scattering reactions with one off-shell 
external leg, working out  real and imaginary parts explicitly. 
We also provide the algorithms for expressing 
them in terms of HPLs and 2dHPLs whose arguments 
lie within the range covered by the numerical routine of~\cite{2dhpl}, 
which allows them to be evaluated numerically. 
Given that the matrix elements are 
linear combinations of the master integrals, this will in turn allow us 
to determine the matrix element for all $2\to 2$ reactions 
related to $e^+e^-\to 3$~jets by crossing. 

In the context of the next-to-leading order (NLO) corrections to 
jet observables, the first calculation~\cite{ert} was also 
for the kinematics of the $1\to 3$ decay, relevant to $e^+e^-\to 3$~jets.
The one-loop matrix element obtained in this calculation contained 
only logarithms and dilogarithms, which have a known analytic continuation. 
The results of~\cite{ert} were continued to the kinematic situation 
relevant to hadronic vector-boson-plus-jet production in~\cite{emp}, 
and to deep inelastic two-plus-one-jet production 
in~\cite{graudenz1,graudenz2}.
The analytic continuation procedure used in these 
calculations is documented in detail in~\cite{graudenz2}. In particular 
it is already observed at the one-loop level that the 
kinematic region relevant to the  hadronic vector-boson-plus-jet production
is free of kinematic cuts, while two kinematic cuts are 
present inside the region of deep inelastic two-plus-one-jet production. 
We shall see below that the same pattern is preserved at the two-loop 
level.

The paper is structured as follows. In 
Section~\ref{sec:kin}, we define the kinematical variables used 
to describe two-loop four-point functions with one off-shell leg and 
describe the ranges of variables relevant to each different process by  
making a detailed decomposition of the kinematic plane. The basic
features of analytic continuation are discussed in Section~\ref{sec:bascont}
on the example of the continuation of the $1\to 3$ decay from the 
Euclidean to the Minkowskian region. 
In Sections~\ref{sec:twocont} and~\ref{sec:onecont}, 
we derive the algorithms for the analytic continuation in one invariant 
for a time-like and a space-like off-shell leg.
Section~\ref{sec:conc}
contains conclusions and an outlook. 
Finally, the Appendix recalls definitions and main properties of the 
HPLs and 2dHPLs. 

\section{Kinematic regions and notation}
\setcounter{equation}{0}
\label{sec:kin}

\begin{figure}[b] 
\begin{center}
\epsfig{file=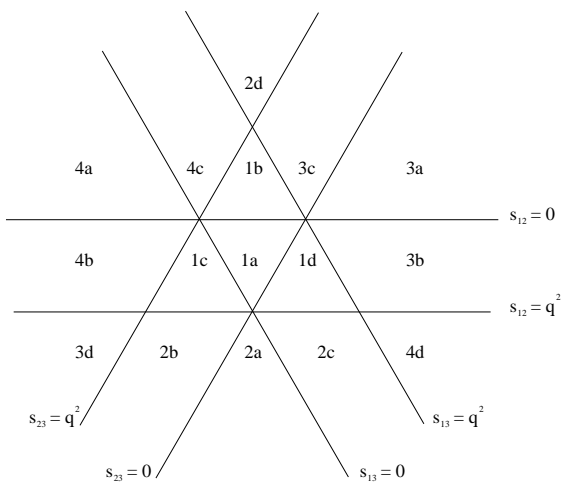,angle=-90,width=9cm}
\end{center}
\caption{Kinematic plane in terms of Lorentz invariants 
$s_{ij} = (p_i+p_j)^2$ displayed in equilateral coordinates.}
\label{fig:master}
\end{figure}

We label the external momenta on the four-point functions with 
$p_1$, $p_2$, $p_3$ and $p_4$, and write momentum conservation as 
$ p_1 + p_2 + p_3 + p_4 = 0\ ,$ with on-shell conditions $p_i^2 = 0$ for 
$i=1,2,3$ while $p_4$ is off-shell $p_4^2 = (-p_1-p_2-p_3)^2 = q^2 \neq 0$. 
We further define 
\begin{equation}
s_{12} = (p_1+p_2)^2\;,\quad s_{13} = (p_1+p_3)^2\;, \quad 
s_{23} = (p_2+p_3)^2\; ,
\end{equation}
so that the Mandelstam relation reads 
\begin{equation} 
s_{12}+ s_{13} + s_{23}= q^2 \ . 
\label{eq:Mand} 
\end{equation} 
We will use the metric in which time-like invariants are positive. 
The kinematic plane defined by $s_{12}$, $s_{13}$ and  $s_{23}$ is shown in 
Fig.~\ref{fig:master}, where equilateral (non-Cartesian) coordinates were 
used to display the symmetry in the three invariants. The lines indicate 
the locations of potential cuts in the four-point functions. For later use,
all regions defined by these cuts are labelled as (1a), (1b), $\ldots$, (4d). 

\begin{figure}[t] 
\begin{center}
\epsfig{file=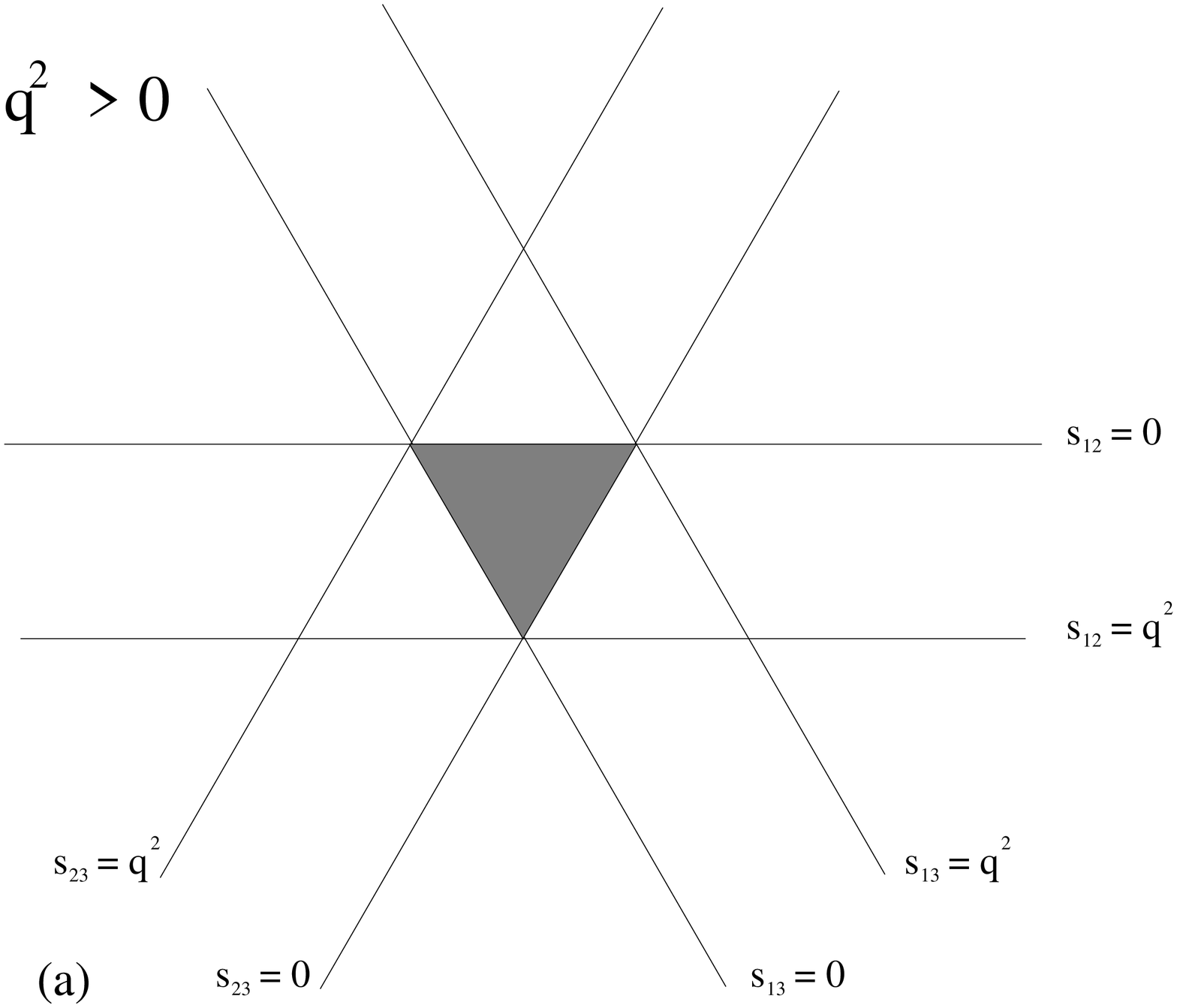,angle=-90,width=8cm}\\
\parbox{16.2cm}{
\epsfig{file=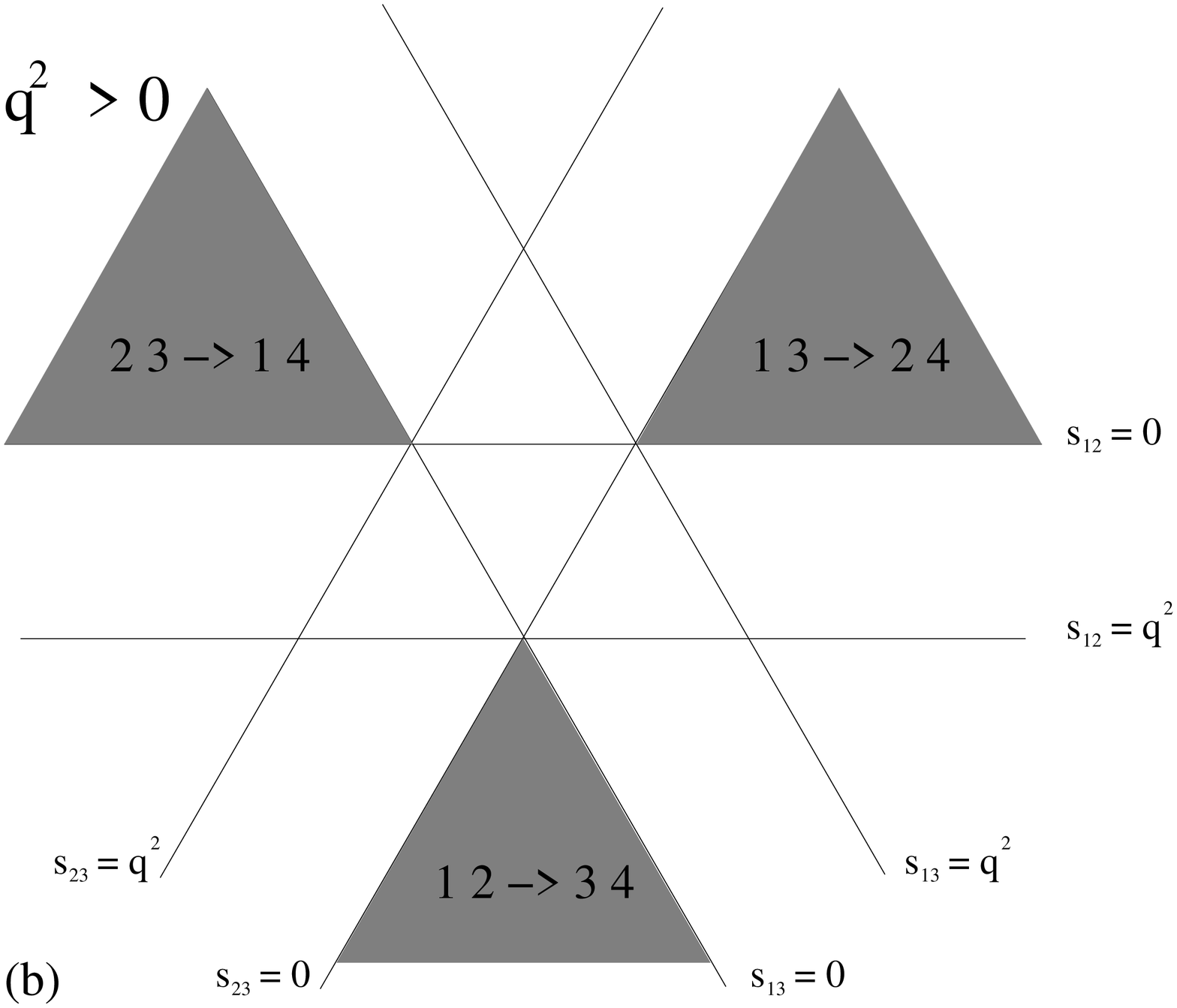,angle=-90,width=8cm}
\epsfig{file=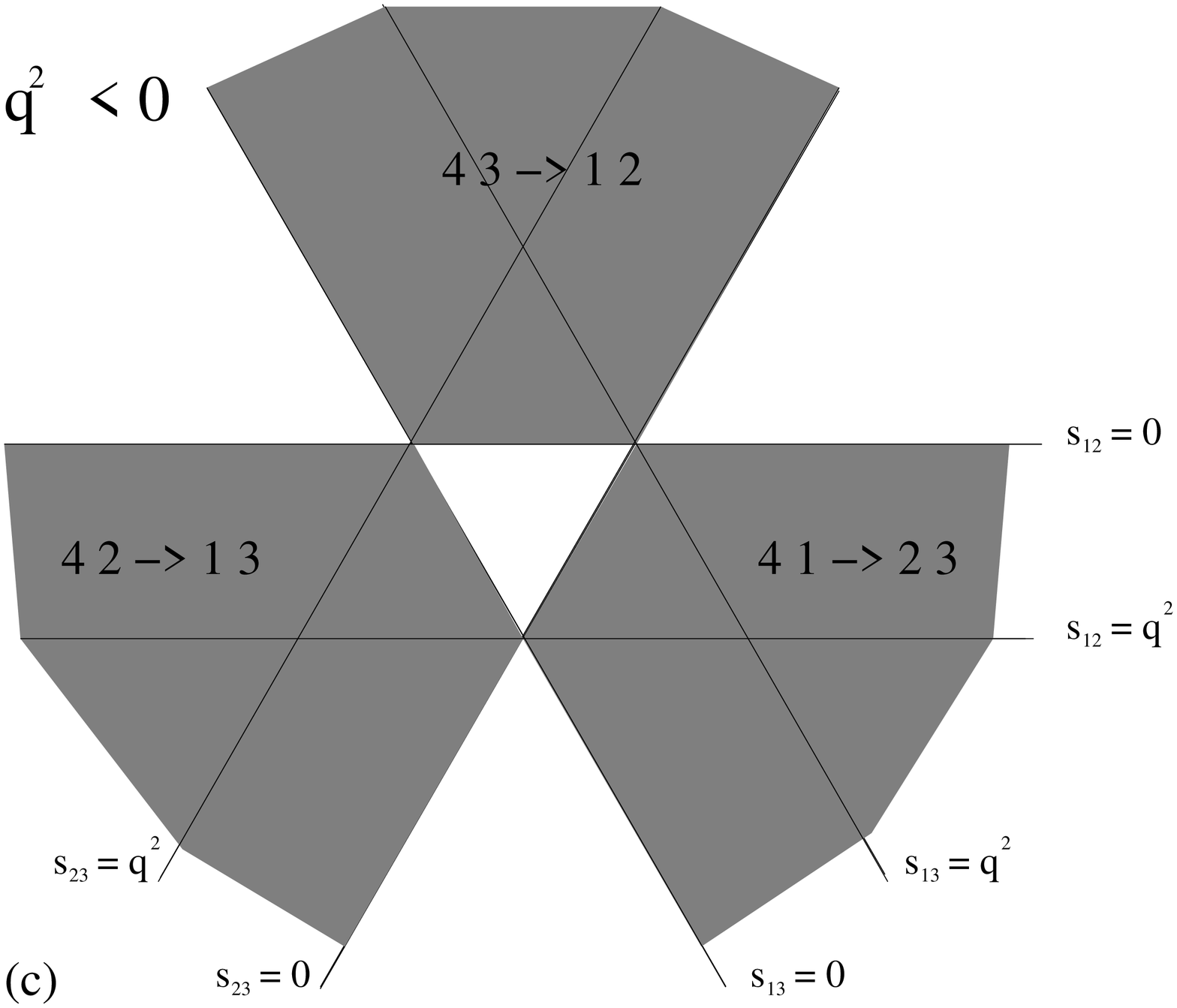,angle=-90,width=8cm}}
\end{center}
\caption{Regions of the kinematic plane relevant to (a) 
$e^+e^- \to 3$~jets $(q^2>0)$, 
(b) $(V+ 1)$~jet production $(q^2>0)$ and 
(c) deep inelastic $(2+1)$~jet production $(q^2<0)$.}
\label{fig:physregions}
\end{figure}

The regions physically relevant to different processes are 
displayed in Fig.~\ref{fig:physregions}.
In $e^+ e^- \to 3$~jet ($3j$) production, $q^2$ is time-like (hence 
positive) and all $s_{ij}$ are positive as well. The relevant region 
is the inner triangle of the kinematic plane, as shown 
in Fig.~\ref{fig:physregions}(a). This inner triangle corresponds to region 
(1a) in Fig.~\ref{fig:master}. To indicate space-like ($q^2<0$) and 
time-like ($q^2>0$) kinematics for a region under discussion, we 
will use in the following 
the subscripts ``$-$'' (space-like) and ``$+$'' (time-like). 
The region relevant to $3j$-production is thus denoted by (1a)$_+$. 

Vector-boson-plus-jet ($V+1j$) production at hadron colliders and 
deep-inelastic two-plus-one-jet (DIS-$(2+1)j$) production are described 
by three subprocesses each (corresponding to the 
$s_{12},s_{13},s_{23}$ channels, which all contribute to these final states). 

For $V+1j$ production, $q^2$ is time-like, and 
for $p_i + p_j \to p_k + p_4$ the invariants fulfil
\begin{equation}
q^2 > 0\;,\quad s_{ij} > q^2 > 0\;,\quad s_{jk} < 0\;, \quad s_{ik} < 0\; ,
\end{equation}
where $(i,j,k)$ stand for the three non-ordered permutations of 
$(1,2,3)$. 
The relevant regions 
are shown in Fig.~\ref{fig:physregions}(b) and correspond to the regions 
(2a,3a,4a)$_+$ of the kinematic plane, Fig.~\ref{fig:master}. 

Finally, for DIS-$(2+1)j$ production, $q^2$ is space-like (hence negative) 
and for $p_4 + p_k \to p_i+p_j$ the invariants fulfil 
\begin{equation}
q^2 < 0\;,\quad s_{ij} > -q^2 > 0\;,\quad s_{jk} < 0, \quad s_{ik} < 0\; ,
\end{equation}
where $(i,j,k)$ stand again for the three non-ordered permutations of 
$(1,2,3)$. We display the relevant kinematic regions in  
Fig.~\ref{fig:physregions}(c). Each of those regions cannot be identified 
with a single region in the kinematic plane, Fig.~\ref{fig:master}, 
but is instead patched together from four regions, (1d,2c,3b,4d), 
(1b,2d,3c,4c) and (1c,2b,3d,4b). 
\begin{figure}[t] 
\begin{center}
\epsfig{file=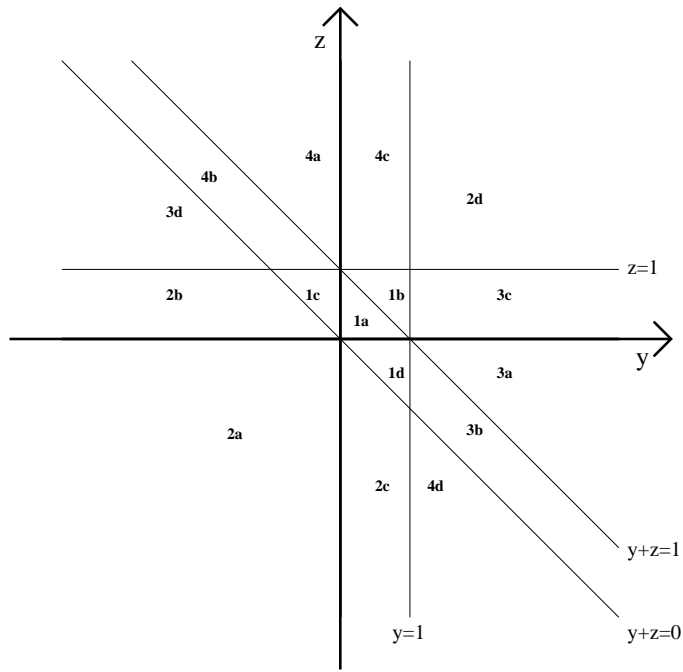,angle=-90,width=9cm}
\end{center}
\caption{Kinematic plane in Cartesian coordinates.}
\label{fig:master_yz}
\end{figure}

It is customary to introduce the dimensionless variables 
\begin{equation}
   x = \frac{s_{12}}{q^2}\;,\quad  y = \frac{s_{13}}{q^2}\; , \quad 
   z = \frac{s_{23}}{q^2}\; ,
\label{eq:defxyz} 
\end{equation}
 for which 
(\ref{eq:Mand}) reads 
\begin{equation} 
x + y + z = 1 \;.
\label{eq:dlMand} 
\end{equation} 
As independent variables, we will use mainly $q^2$, $y$ and $z$, with $x$ 
therefore given as $x=1-y-z$. We will further represent the various 
kinematical configurations, at given $q^2$, in the Cartesian $y,z$ plane. 
For convenience of later use, we represent in 
Fig.~\ref{fig:master_yz} the whole $y,z$ plane properly partitioned in 
all the regions which will be of interest later. The labelling of the regions 
is as in Fig.~\ref{fig:master}.

When the three regions ($3j$, $V+1j$, DIS-$(2+1)j$), 
are superimposed, regardless of $q^2$, they cover the entire 
$(y,z)$ plane.  

Despite using mainly $q^2$, $y$ and $z$ as independent variables, 
all analytic continuations are to be carried out in the 
original Mandelstam variables 
$s_{12}$, $s_{13}$ and $s_{23}$, adding to them, when time-like, the 
usual infinitesimal imaginary part with positive sign, 
$s_{ij} + \iep$.  
Indeed, no definite sign can be attributed {\it a priori} to the imaginary 
parts of the dimensionless invariants $x$, $y$ and $z$, 
given the constraint $x+y+z=1$. In practice, this means that 
any function of $y$ and $z$ that is to be continued analytically, 
has to be expressed first in terms of $s_{12}$, $s_{13}$ and $s_{23}$, 
then continued taking correct account of the well defined imaginary 
parts of the $s_{ij}$, then finally re-expressed in $q^2$ and 
$y,z$ or other dimensionless variables appropriate to the considered region.

\section{Basics of analytic continuation}
\setcounter{equation}{0}
\label{sec:bascont}

The problem of analytic continuation is best discussed starting from 
the unphysical Euclidean case in which $q^2, s_{12}, s_{13}$ and $s_{23}$ 
are all space-like (hence negative). Let us give them the values 
\begin{equation} 
q^2 = - Q^2, \quad s_{ij} = - \sigma_{ij} \end{equation} 
with 
\begin{equation} 
   Q^2>0\;,\quad \sigma_{ij}>0\;,\quad 
   \sigma_{12}+\sigma_{13}+\sigma_{23}=Q^2\;. 
\label{eq:Q2andsigmas} 
\end{equation}

All master integrals are indeed real in this configuration. 
According to Eq.~(\ref{eq:defxyz}), we introduce 
\begin{equation} 
   x = \frac{s_{12}}{q^2} = \frac{\sigma_{12}}{Q^2} \; , \quad 
   y = \frac{s_{13}}{q^2} = \frac{\sigma_{13}}{Q^2} \; , \quad 
   z = \frac{s_{23}}{q^2} = \frac{\sigma_{23}}{Q^2} \; , 
\nonumber \label{eq:sigdefxyz} 
\end{equation} 
and observe that, because of Eq.~(\ref{eq:Q2andsigmas}), we have 
$ (0 < y < 1,\ 0 < z < 1-y) $ or equivalently 
$ (0 < z < 1,\ 0 < y < 1-z )$, i.e.~$(y,z)$ 
are in the region (1a) of Fig.~\ref{fig:master_yz}. 
Any of the master integrals evaluated in~\cite{doublebox}, say 
$\Psi(s_{12},s_{13},s_{23})$, can be written as 
\begin{eqnarray} 
\Psi(s_{12},s_{13},s_{23}) &=& ( Q^2 )^{\alpha} \Phi\left( 
    \frac{\sigma_{13}}{Q^2}, \frac{\sigma_{23}}{Q^2} \right) \nonumber\\ 
         &=& ( - q^2 )^{\alpha} \Phi\left( 
    \frac{s_{13}}{q^2}, \frac{s_{23}}{q^2} \right) \nonumber\\ 
         &=& ( - q^2 )^{\alpha} \Phi(y,z) \ , 
\label{eq:start} 
\end{eqnarray} 
where use is made of Eq.~(\ref{eq:sigdefxyz}) and 
$\alpha$ is an exponent accounting for the mass dimension of the 
integral: $\alpha = d - r + s$, where $d$ is the continuous space-time 
dimension used in dimensional regularization, while the integers $r$ and 
$s$ denote  the number of propagators and scalar products in the 
integral. For a two-loop seven-propagator master integral 
(e.g.\ a scalar double box or crossed box integral), with all 
propagators raised to unit power and no scalar products, one 
thus has $\alpha = d-7 = - 3 +(d-4) $. The function $ \Phi(y,z) $ 
is in general a Laurent series in $(d-4)$, whose coefficients are 
combinations of simple algebraic factors in $y, z$ 
times HPLs of various weight and argument $z$ and 2dHPLs of various weight 
and argument $y$, with indices depending on $z$. 

Note that the dimensionless ratios $y,z$ are both real and positive, 
and lie within the boundaries $0<z<1$, $0<y<1-z$. In that range of values 
of the arguments, all the HPLs and 2dHPLs are analytic and real -- as 
expected, of course, as all the kinematical variables are Euclidean. 
This region coincides with the triangle (1a)$_-$ of Fig.~\ref{fig:master_yz}. 
 
As a consequence of the symmetry and analyticity properties of the 
HPLs and 2dHPLs, within the region (1a)$_-$ we can exchange the roles 
of $y$ and $z$ and re-express the master integrals as 
\begin{equation} 
\Psi(s_{12},s_{13},s_{23}) = ( - q^2 )^{\alpha} \Phi'(z,y) \ , 
\label{eq:startzy} 
\end{equation} 
where $\Phi'(z,y)$ consists of HPLs of argument $y$ and 2dHPLs of argument 
$z$ with indices depending on $y$, all within their analyticity region. 
The transformation from $\Phi(y,z)$ to $\Phi'(z,y)$ 
can be implemented, within the triangle (1a) of Fig.~\ref{fig:master_yz}, 
by re-expressing the HPLs $\H(\vec{m};z)$ and 2dHPLs $\G(\vec{m}(z);y)$  
as a combination of HPLs $\H(\vec{m};y)$ and 2dHPLs $\G(\vec{m}(y);z)$, 
by means of the `interchange-of-arguments' procedure 
described in Appendix A.2.2 of~\cite{doublebox}.
All algorithms used for this procedure (and for all subsequent 
transformations derived in this paper)
were coded in FORM~\cite{form}.

Similarly,  within the region (1a) 
we can replace $z=1-x-y$ by $x$, obtaining another representation 
of the master integrals 
\begin{equation} 
\Psi(s_{12},s_{13},s_{23}) = ( - q^2 )^{\alpha} \Phi''(y,x) \ , 
\label{eq:startyx} 
\end{equation} 
where $\Phi''(y,x)$ consists of HPLs of argument $x$ and 2dHPLs of argument 
$y$, with indices depending on $x$, all within their analyticity region. 
The above transformation, which amounts to re-expressing 
the HPLs $\H(\vec{m};z)$ and 2dHPLs $\G(\vec{m}(z);y)$ 
appearing in the original $\Phi(y,z)$ in terms of 
HPLs $\H(\vec{n};x)$ and 2dHPLs $\G(\vec{n}(x);y)$, with $x=1-y-z$, 
can be implemented 
as a result of the combination of the `interchange-of-arguments'-procedure 
described in Appendix A.2.2 of~\cite{doublebox} with the 
reflection algorithm derived in Section 5 of~\cite{2dhpl}.

The notation for the 2dHPLs used in the present work is shortly 
recalled in the Appendix. It is the same as was introduced in~\cite{2dhpl}, 
which is however different from the notation originally proposed and used 
in~\cite{doublebox} to present the two-loop four-point master integrals. 
The notation of~\cite{2dhpl} was already employed 
in~\cite{3jme} to represent the result for the two-loop QCD 
corrections to the $3j$ matrix element.
Detailed transformation 
rules between the two notations can be found in~\cite{2dhpl}. 
A summary of 
definitions and  properties of HPLs and 2dHPLs is provided in the 
appendix of~\cite{3jme}. 

In the $3j$ case, all the kinematical variables are time-like, hence 
positive; the proper analytic continuation is obtained by starting 
from the Euclidean case and giving to each positive variable (hence 
to all the variables in this case) a small positive imaginary 
part (the usual $+\iep$ prescription). 
Equation (\ref{eq:start}) then becomes 
\begin{equation} 
\Psi(s_{12}+\iep,s_{13}+\iep,s_{23}+\iep) = ( - q^2 - \iep )^{\alpha} 
   \Phi\left( \frac{s_{13}+\iep}{q^2+\iep}, \frac{s_{23}+\iep}{q^2+\iep} 
                                            \right). 
\label{eq:3j1} 
\end{equation} 
The real parts of $(s_{13}+\iep)/(q^2+\iep)$ and $(s_{23}+\iep)/(q^2+\iep)$, 
\begin{equation}
\frac{s_{13}}{q^2} = y \;,\quad\quad \frac{s_{23}}{q^2} = z \;,
\end{equation}
lie in the region 
$0<y<1$, $0<z<1-y$, the triangle (1a) of Fig.~\ref{fig:master_yz}, where 
all the HPLs and 2dHPLs are analytic and real. The $\epsilon \to 0$ 
limit is then trivial (i.e.\ the $\iep$ can be simply ignored) and 
Eq.~(\ref{eq:3j1}) becomes 
\begin{equation} 
\Psi(s_{12}+\iep,s_{13}+\iep,s_{23}+\iep) = ( - q^2 - \iep )^{\alpha} 
            \Phi(y,z) \ , 
\label{eq:3j2} 
\end{equation} 
where $ \Phi(y,z) $ is exactly the same as in the fully Euclidean case. 
The continuation from (1a)$_-$ to (1a)$_+$ performed here 
leaves  $ \Phi(y,z) $ unchanged.
The imaginary parts of the $3j$ master integrals (which are of course 
complex) are entirely due to the factor $( - q^2 - \iep )^{\alpha}$. 
The integer part of $\alpha$ does not matter, while the expansion in 
$(d-4)$ gives 
\begin{equation}
( - q^2 - \iep )^{d-4} = 1 + (d-4)\bigl(\ln(q^2) - i\pi\bigr) 
             + \frac{1}{2}(d-4)^2 \bigl(\ln(q^2) - i\pi\bigr)^2 + ... \;.
\end{equation}
It is worth recalling that the master integrals, as well as the physical
matrix elements, develop polar singularities around $d=4$ in dimensional 
regularization, so that these higher-order terms in the 
$(d-4)$-expansion and their imaginary parts become of actual importance. 

It is to be emphasised here that in~\cite{doublebox}, strictly speaking, 
the master integrals $ \Psi(s_{12},s_{13},s_{23}) $ were never directly 
evaluated in the fully Euclidean region; but as $\Phi(y,z)$ is the same 
function, and for the same range of arguments, in both the fully Euclidean 
case, Eq.~(\ref{eq:start}), and the $3j$ case, 
Eq.~(\ref{eq:3j2}), we can define in that way (i.e.\ by just replacing in 
the overall scale factor $(-q^2-\iep)^\alpha$ by 
$(Q^2)^\alpha$) the Euclidean master 
integrals in terms of the master integrals given in \cite{doublebox} for 
the $3j$ case. From now on, we will therefore take the Euclidean 
master integrals as known, and will 
show how to get by analytic continuation the master integrals in the 
various kinematical regions of physical interest.  

\section{Analytic continuation in one invariant for time-like $q^2$}
\setcounter{equation}{0}
\label{sec:twocont}

To continue from region (1a)$_-$ to regions (2a,3a,4a)$_+$, 
which are relevant to
 $V+1j$ production
at hadron colliders (with a time-like momentum of the 
vector boson $V$),
it is necessary to continue simultaneously in one of three Lorentz
invariants $s_{ij}$ and in the vector boson virtuality $q^2$. \par 
We start by discussing the continuation from (1a)$_-$ to (4a)$_+$
\footnote{Notice that the analytic continuation from (1a)$_-$ to 
(3a)$_+$ as 
outlined in the appendix of~\cite{doublebox} is unnecessarily complicated. 
Moreover, Eq.~(A.30) in~\cite{doublebox}, which is relevant in this context, 
contains a misprint: all $i$'s should read $-i$.}. 

In region (4a)$_+$ of Fig.~\ref{fig:master_yz}, relevant for $V+1j$ 
production in the momentum arrangement $p_2+p_3 \to p_1+p_4$, 
($1<z<\infty,\ 1-z<y<0$) or ($-\infty<y<0,\ 1-y<z<\infty$) and 
\begin{equation}
q^2 > 0,\ \ s_{12} < 0,\ \ s_{13} < 0,\ \ \ 
s_{23} = q^2 - s_{12} - s_{13} > q^2 > 0 \ , 
\label{eq:4a+1} 
\end{equation} 
so that for the analytic continuation $q^2$ and $ s_{23} $ must be given 
infinitesimal imaginary parts $+\iep$. 
From Eq.~(\ref{eq:start}) one gets in this case 
\begin{equation}
\Psi(s_{12},s_{13},s_{23}+\iep) = (-q^2-\iep)^\alpha \Phi\left( 
     \frac{s_{13}}{q^2+\iep}, \frac{s_{23}+\iep}{q^2+\iep} \right) 
   = (-q^2-\iep)^\alpha \Phi(y+\iep,z-\iep) , 
\label{eq:start4aplus} 
\end{equation} 
where, because of Eqs.~(\ref{eq:4a+1}) and of the very definition of $y,z$, 
Eqs.~(\ref{eq:defxyz}), we have used 
\begin{equation}
 \frac{s_{13}}{q^2+\iep} = \frac{s_{13}}{q^2} +\iep = y +\iep \ ,\ \ \ 
 \frac{s_{23}+\iep}{q^2+\iep} = \frac{s_{23}}{q^2} -\iep = z -\iep \ . 
\label{eq:4a+3} 
\end{equation} 
We introduce new dimensionless variables $u, v$ through the relations 
\begin{equation}
u_{{\rm 4a}} =u  = - \frac{s_{13}}{s_{23}} = - \frac{y}{z}, \qquad 
v_{{\rm 4a}} =v  =   \frac{q^2}{s_{23}} = \frac{1}{z} \; ,
\label{eq:4a+4} 
\end{equation} 
so that 
\begin{equation}
 y = - \frac{u}{v} \ ,\ \ \ z = \frac{1}{v} \ . 
\label{eq:start4a}  
\end{equation} 
As $y$ and $z$ span the region (4a) 
of Fig.~\ref{fig:master_yz}, we find that $u, v$ vary in the ranges 
$ 0 < v < 1 $ and $ 0<u<1-v$, \ie\ the 
above parametrisation maps region (4a) into region (1a). 
\par 
Equation (\ref{eq:start4aplus}) then reads 
\begin{equation}
\Psi(s_{12},s_{13},s_{23}+\iep) = (-q^2-\iep)^\alpha \Phi\left( 
         - \frac{u}{v} + \iep, \frac{1}{v} - \iep\right) . 
\label{eq:start4aplus1} 
\end{equation}

Given the above prescription, it is relatively straightforward to 
determine the proper analytic continuation of the various HPLs and 2dHPLs 
appearing in the analytic expression $ \Phi(y,z) $ of any master integral 
and then to express them as HPLs of argument $v$ and 2dHPLs of argument 
$u$ and indices depending on $v$, with $(u, v)$, as already observed, 
within the analyticity triangle of the functions. 
Let us start from the HPLs and 2dHPLs of weight $w$ equal to 1, which 
are just logarithms. According to the definitions of the appendix, one has 
\begin{eqnarray}
\H(0;z-\iep) & = & \log\left(\frac{1}{v}-\iep\right) = -\log{v}
             \nonumber \\ 
& = & -\H(0;v)\;, \nonumber \\ 
\H(1;z-\iep) & = & - \log\left(1 - \frac{1}{v} + \iep \right) 
          = - \log(1-v) + \log{v} - i\pi \nonumber \\ 
& = & \H(1;v) + \H(0;v) - i\pi \; .
\label{eq:hcona}
\end{eqnarray}
Notice that for $\H(0;z)$, the $\iep$-term can safely be ignored. 

The 2dHPLs at $w=1$, again according to the definitions of the appendix, 
are continued as
\begin{eqnarray}
\G(0;y+\iep) & = & \log\left(-\frac{u}{v}+\iep\right) 
= \log\left(\frac{u}{v}\right) + i\pi\nonumber \\ 
& = & \G(0;u) - \H(0;v) + i\pi \nonumber \\
\G(1;y+\iep) & = & \log\left(1+\frac{u}{v}-\iep\right) \nonumber \\
& = & \G(-v;u) \nonumber \\
\G(1-z+\iep;y+\iep) & = & \log\left(\frac{1+\frac{u}{v}-\iep-\frac{1}{v}+\iep}
{1-\frac{1}{v}+\iep}\right) = \log\left(\frac{1-u-v}
{1-v}\right)\nonumber \\
& = & \G(1-v;u) \nonumber\\
\G(-z+\iep;y+\iep) & = & \log\left(\frac{-\frac{u}{v}+\iep+\frac{1}{v}-\iep}
{\frac{1}{v}-\iep}\right) = \log (1-u) \nonumber \\
& = & \G(1;u)\; .
\label{eq:gcona}
\end{eqnarray} 
It should be noted that no definite imaginary parts can be assigned to 
the arguments of the logarithms in $\G(1-z+\iep;y+\iep)$ and 
$\G(-z+\iep;y+\iep)$; but as 
the arguments in both cases remain positive and within the analyticity 
region of the functions, the result is anyhow well defined. 

Using the above formulae, all imaginary parts of the higher weight 
HPLs and 2dHPLs are fixed, since these functions can be derived iteratively, 
starting from the $w=1$ functions, as will now be shown. 

If, in an HPL of weight $w$ higher than $1$, all the indices are equal 
to $1$, that HPL is just equal to $1/w!$ times the $w$-th power of 
$\H(1;z-\iep)$, already seen. When the indices 
are not all equal to 1, one can first 
separate all leftmost $(1)$'s in the index vector, using the product 
algebra (\ref{eq:halgebra}), such that the leftmost 
index is always a $(0)$. Any HPL can then be written as~\cite{1dhpl} 
\begin{eqnarray}
\H(0,\vec{b};z-\iep) & = & \H(0,\vec{b};1-\iep) + \int_{1-\iep}^{z-\iep}\d z' 
                         \frac{\d}{\d z'} \H(0,\vec{b};z') \nonumber \\ 
                     & = & \H(0,\vec{b};1) + \int_1^{z}\d z' 
                         \frac{\d}{\d z'} \H(0,\vec{b};z'-\iep) \nonumber \\ 
                     & = & \H(0,\vec{b};1) - \int_1^v \frac{\d v'}{v'} 
                           \H\left(\vec{b};\frac{1}{v'}-\iep\right) \; ,
\end{eqnarray} 
where in the last step the integration variable $v'=1/z'$ was introduced. 
The expression for $\H(\vec{b};1/v'-\iep)$ in terms of 
$\H(\vec{c};v')$ and its proper imaginary part 
is of lower weight than $\H(0,\vec{b};z-\iep)$ and thus already known in 
an iterative bottom up approach in the weight $w$. 
As an example of this transformation, one finds
\begin{eqnarray}
\H(0,1,1;z-\iep) & = & \H(0,0,1;v) - \H(0,1,1;v) 
                + \H(0;v)\biggl( \frac{\pi^2}{2} - \H(0,1;v) - \frac{1}{6}
         \H(0;v)\H(0;v) \biggr) + \zeta_3 \nonumber \\
           && - i\pi \biggl(- \H(0,1;v) - \frac{1}{2}
         \H(0;v)\H(0;v) + \frac{\pi^2}{6}  \biggr)\ .
\end{eqnarray}

To perform the analytic continuation of the 2dHPL in $y$, one first separates 
off all rightmost $(0)$'s in the index vector by applying the product algebra.
The remaining 2dHPL (the dependence of the indices $a, \vec{b}$ on 
$z=1/v-\iep$ is understood for short) can then be written as 
\begin{eqnarray}
\G\left(a,\vec{b};-\frac{u}{v}+\iep\right) & = & 
                \int_0^{-\frac{u}{v}+\iep} \d y' \frac{\d}{\d y'} 
\G(a,\vec{b};y')\nonumber \\
 & = & \int_0^{-\frac{u}{v}} \d y' \frac{\d}{\d y'} 
\G(a,\vec{b};y'+\iep)\nonumber \\
& = & \int_0^{-\frac{u}{v}} \d y' \g(a,y'+\iep) \G(\vec{b};y'+\iep) 
                                                                \nonumber \\
& = &  - \int_0^u \d u'\ \frac{1}{v} 
       \g\left(a,-\frac{u'}{v}+\iep\right)\, 
       \G\left(\vec{b};-\frac{u'}{v}+\iep\right),
\label{eq:ggen}
\end{eqnarray}
where the expression for $\G(\vec{b};-u'/v+\iep)$ in terms of 
$\G(\vec{c};u')$ and its proper imaginary part 
is again of lower weight and thus known in an iterative bottom up 
approach in $w$. In the rational fractions the $\iep$ does not matter 
and their expressions are
\begin{eqnarray}
\frac{1}{v} \g\left(0;-\frac{u'}{v}  \right) &=& -\g(0;u')\; ,\nonumber \\
\frac{1}{v} \g\left(1;-\frac{u'}{v}  \right) &=& -\g(-v;u')\; , \nonumber \\
\frac{1}{v} \g\left(1-z;-\frac{u'}{v}\right) &=& -\g(1-v;u')\;, \nonumber \\
\frac{1}{v} \g\left(-z;-\frac{u'}{v} \right) &=& -\g(1;u') \; , 
\end{eqnarray} 
such that the $u'$-integral  in (\ref{eq:ggen}) yields a 2dHPL of argument $u$.

From the above, it becomes immediately clear that only a 2dHPL with trailing 
$(0)$'s in the index vector acquire an imaginary part when continued from 
(1a)$_-$ to (4a)$_+$. Moreover, any 2dHPL from (1a)$_-$
without trailing $(0)$'s is identified 
with a single 2dHPL in (4a)$_+$ (since the $w=1$ functions (\ref{eq:gcona})
are identified on a one-to-one basis). 

An example of the continuation of a 2dHPL is 
\begin{equation}
\G(0,1,1-z,0;y+\iep) = 
     - \H(0,v) \; \G(0,-v,1-v,u) 
     + \G(0,-v,1-v,0,u)
     + i \pi \; \G(0,-v,1-v;u) \;. 
\label{eq:exofcont} 
\end{equation} 

To summarize, master integrals and matrix elements, which are given in 
(1a)$_-$ in terms of HPLs $\H(\vec{m};z)$ and 2dHPLs $\G(\vec{m}(z);y)$ 
can be continued to (4a)$_+$, by using Eq.~(\ref{eq:start4a}), where they are 
expressed in terms of 
HPLs $\H(\vec{m};v)$ and  2dHPLs $\G(\vec{m}(v);u)$. Given the definitions 
$u=-y/z$ and $v=1/z$, one finds $0\leq v \leq 1$, $0\leq u\leq 1-v$ in (4a),
such that the above HPLs and 2dHPLs are real, and can be numerically 
evaluated with the routines of~\cite{1dhpl,2dhpl}. Imaginary parts were made 
explicit in the analytic continuation.

The other two momentum arrangements relevant to vector boson 
production at hadron colliders are $p_1+p_2 \to p_3 + p_4$, corresponding 
to region (2a)$_+$ and $p_1 + p_3 \to p_2 + p_4$, 
corresponding to region (3a)$_+$. 

In the region (2a)$_+$, we have ($-\infty<z<1,\ -\infty<y<1$) and  
\begin{equation}
q^2 > 0,\ \ s_{12} = q^2 - s_{13} - s_{23} > q^2 > 0 ,\ \ s_{13} < 0,\ \ \ 
s_{23} < 0 \ , 
\label{eq:2a+1} 
\end{equation} 
so that $q^2$ and $ s_{12} $ must, for the analytic continuation, be given 
infinitesimal imaginary parts $+\iep$. 
\par 
In close analogy with the discussion for the region (4a)$_+$, we start 
from Eq.~(\ref{eq:start}), which now becomes 
\begin{equation}
\Psi(s_{12}+\iep,s_{13},s_{23}) = (-q^2-\iep)^\alpha \Phi\left( 
     \frac{s_{13}}{q^2+\iep}, \frac{s_{23}}{q^2+\iep} \right) 
  = (-q^2-\iep)^\alpha \Phi(y+\iep,z+\iep) \ , 
\label{eq:master2aplus} 
\end{equation} 
with 
\begin{equation}
 \frac{s_{13}}{q^2+\iep} = \frac{s_{13}}{q^2} +\iep = y +\iep \ ,\ \ \ 
 \frac{s_{23}}{q^2+\iep} = \frac{s_{23}}{q^2} +\iep = z +\iep \ ; 
\label{eq:2a+3} 
\end{equation} 
we then introduce new dimensionless variables $u_{{\rm{2a}}}, v_{{\rm{2a}}}$ 
as 
\begin{equation}
u_{{\rm 2a}} =  - \frac{s_{13}}{s_{12}} = - \frac{y}{1-y-z}, \qquad 
v_{{\rm 2a}} =   \frac{q^2}{s_{12}} = \frac{1}{1-y-z} \; ,
\label{eq:2a+4} 
\end{equation} 
so that 
\begin{equation}
 y = - \frac{u_{{\rm 2a}}}{v_{{\rm 2a}}} \ , 
 \ \ \ z = - \frac{1-u_{{\rm 2a}}-v_{{\rm 2a}}}{v_{{\rm 2a}}} \ ,  
\label{eq:start2a}  
\end{equation} 
and $u_{{\rm 2a}}, v_{{\rm 2a}}$ vary in the ranges 
$ 0 < v_{{\rm 2a}} < 1 $ and $ 0<u_{{\rm 2a}}<1-v_{{\rm 2a}}$, \ie\ the 
above parametrization maps region (2a) into region (1a), 
and Eq.~(\ref{eq:master2aplus}) becomes 
 \begin{equation}
\Psi(s_{12},s_{13},s_{23}+\iep) = (-q^2-\iep)^\alpha \Phi\left( 
         - \frac{u_{{\rm 2a}}}{v_{{\rm 2a}}} + \iep, 
         - \;\frac{1-u_{{\rm 2a}}-v_{{\rm 2a}}}{v_{{\rm 2a}}} + \iep\right) . 
\end{equation} 
The separation of the real and imaginary parts of the HPLs and 2dHPLs 
of the above arguments 
$-u_{{\rm 2a}}/v_{{\rm 2a}} + \iep , 
(1-u_{{\rm 2a}}-v_{{\rm 2a}})/v_{{\rm 2a}} + \iep $, 
and their expression in terms of 
HPLs and 2dHPLs of arguments $u_{{\rm 2a}},v_{{\rm 2a}}$ and $(i\pi)$'s 
can then be carried out by a suitable extension of the derivation of 
Eqs.~(\ref{eq:hcona})--(\ref{eq:exofcont}). 
\par 
Alternatively, we can use the representation (\ref{eq:startyx}) of 
the master integrals, \ie\ we can first transform the expression for 
the Euclidean master integral (\ref{eq:start}) by rewriting $\Phi(y,z)$ 
in terms of $x=1-y-z$, so obtaining the function $\Phi''(y,x)$ defined as 
\begin{equation}
\Phi''(y,x) = \Phi(y,1-x-y) \; . 
\end{equation}

For the continuation to (2a)$_+$, Eq.~(\ref{eq:startyx}) reads 
\begin{equation}
\Psi(s_{12}+\iep,s_{13},s_{23}) = (-q^2-\iep)^\alpha \Phi''\left( 
     \frac{s_{13}}{q^2+\iep}, \frac{s_{12}+\iep}{q^2+\iep} \right) 
  = (-q^2-\iep)^\alpha \Phi''(y+\iep,x-\iep) , 
\label{eq:master2a} 
\end{equation} 
with 
\begin{equation}
 \frac{s_{13}}{q^2+\iep} = \frac{s_{13}}{q^2} +\iep = y +\iep \ ,\ \ \ 
 \frac{s_{12}+\iep}{q^2+\iep} = \frac{s_{12}}{q^2} -\iep = x -\iep \ . 
\end{equation} 
The dimensionless variables $u_{{\rm{2a}}}, v_{{\rm{2a}}}$ of 
Eq.~(\ref{eq:2a+4}) can also be written as 
\begin{equation}
u_{{\rm 2a}} =  - \frac{s_{13}}{s_{12}} = - \frac{y}{x}, \qquad 
v_{{\rm 2a}} =   \frac{q^2}{s_{12}} = \frac{1}{x} \; ,
\end{equation} 
so that 
\begin{equation}
 y = - \frac{u_{{\rm 2a}}}{v_{{\rm 2a}}} \ , 
 \ \ \ x = \frac{1}{v_{{\rm 2a}}} \ ,  
\end{equation} 
and Eq.~(\ref{eq:master2a}) becomes 
\begin{equation}
\Psi(s_{12}+\iep,s_{13},s_{23}) = (-q^2-\iep)^\alpha \Phi''\left( 
         - \frac{u_{{\rm 2a}}}{v_{{\rm 2a}}} + \iep, 
          \;\frac{1}{v_{{\rm 2a}}} - \iep\right) \ . 
\end{equation} 
The same analytic continuation formulae as applied above for 
the continuation of Eq.~(\ref{eq:start4aplus1}) to (4a)$_+$ can therefore 
be used in this case as well, if allowance 
is made for the formal replacement $z\to x$ and $v\to v_{{\rm 2a}}$. 

In the region (3a)$_+$ , corresponding to $p_1 + p_3 \to p_2 + p_4$, 
($ -\infty<z<0,\ 1-z<y<\infty$) and 
\begin{equation}
q^2 > 0,\ \ s_{12} < 0 \ , \ \ s_{13}= q^2 - s_{12} - s_{23} > q^2 > 0 \ , 
 \ \ \ \ s_{23} < 0 \ , 
\label{eq:3a+1} 
\end{equation} 
for the analytic continuation $q^2$ and $ s_{13} $ must be given 
imaginary parts $+\iep$, 
\begin{equation}
\Psi(s_{12},s_{13}+\iep,s_{23}) = (-q^2-\iep)^\alpha \Phi\left( 
     \frac{s_{13}+\iep}{q^2+\iep}, \frac{s_{23}}{q^2+\iep} \right) 
    = (-q^2-\iep)^\alpha \Phi(y-\iep,z+\iep) \ . 
\end{equation} 
We introduce new dimensionless variables $u_{{\rm{3a}}}, v_{{\rm{3a}}}$ as 
\begin{equation}
u_{{\rm 3a}} =  - \frac{s_{23}}{s_{13}} = - \frac{z}{y}, \qquad 
v_{{\rm 3a}} =   \frac{q^2}{s_{13}} = \frac{1}{y} \; ,
\label{eq:3a+4} 
\end{equation} 
so that 
\begin{equation}
 y = \frac{1}{v_{{\rm 3a}}} \ , 
 \ \ \ z = - \frac{u_{{\rm 3a}}}{v_{{\rm 3a}}} \ ,  
\label{eq:start3a}  
\end{equation} 
$u_{{\rm 3a}}, v_{{\rm 3a}}$ vary in the ranges 
$ 0 < v_{{\rm 3a}} < 1 $ and $ 0<u_{{\rm 3a}}<1-v_{{\rm 3a}}$, \ie\ the 
above parametrization maps region (3a) into region (1a), and the 
proper analytic continuation is given by 
\begin{equation}
\Psi(s_{12},s_{13}+\iep,s_{23}) = (-q^2-\iep)^\alpha \Phi\left( 
           \frac{1}{v_{{\rm 3a}}} - \iep, 
         - \;\frac{u_{{\rm 3a}}}{v_{{\rm 3a}}} + \iep\right) . 
\end{equation} 

In close analogy with the (2a)$_+$ case, the 
separation of the real and imaginary parts of the HPLs and 2dHPLs 
of arguments $1/v_{{\rm 3a}} - \iep , - u_{{\rm 3a}}/v_{{\rm 3a}} + \iep $ 
and their expression in terms of 
HPLs and 2dHPLs of arguments $u_{{\rm 3a}},v_{{\rm 3a}}$ and $(i\pi)$'s 
can then be carried out by a suitable extension of the 
derivation of Eqs.~(\ref{eq:hcona})--(\ref{eq:exofcont}) previously 
established for the region (4a)$_+$. 
\par 

Alternatively, we can start from the expression Eq.~(\ref{eq:startzy}) for 
the Euclidean master integral obtained by crossing the arguments:
\begin{eqnarray} 
\Psi(s_{12},s_{13},s_{23}) &=&  ( - q^2 )^{\alpha} \Phi(y,z) \nonumber\\
&=& ( - q^2 )^{\alpha} \Phi'(z,y) 
\label{eq:master3a} 
\\
&=& ( - q^2 )^{\alpha} \Phi'\left(\frac{s_{23}}{q^2}
,\frac{s_{13}}{q^2}\right)\;.
 \nonumber
\end{eqnarray} 

The expression (\ref{eq:master3a}) then reads 
\begin{eqnarray} 
\Psi(s_{12},s_{13}+\iep,s_{23}) &=& (-q^2-\iep)^\alpha \Phi'\left( 
     \frac{s_{23}}{q^2+\iep}, \frac{s_{13}+\iep}{q^2+\iep} \right) 
     \nonumber\\ 
   &=& (-q^2-\iep)^\alpha \Phi'(z+\iep,y-\iep) \nonumber\\ 
   &=& (-q^2-\iep)^\alpha \Phi'\left( 
     - \;\frac{u_{{\rm 3a}}}{v_{{\rm 3a}}} + \iep, 
      \frac{1}{v_{{\rm 3a}}} - \iep\right) \ , 
\end{eqnarray} 
with $u_{{\rm 3a}},v_{{\rm 3a}}$ given by Eq.~(\ref{eq:3a+4}). 
To this expression, we can again apply the same analytic continuation 
formulae as above, when allowance is made of the formal replacement 
$y\leftrightarrow z$ and $u\to u_{{\rm 2a}}$, $v\to v_{{\rm 2a}}$.

\section{Analytic continuation in one invariant for space-like $q^2$} 
\setcounter{equation}{0} 
\label{sec:onecont} 

To continue from region (1a)$_-$  to regions (1bcd,2bcd,3bcd,4bcd)$_-$, 
which are relevant to deep inelastic two-plus-one-jet production,
it is necessary to continue one of the three Lorentz invariants $s_{ij}$ 
to the time-like region, while not altering the negative sign of $q^2$. 

In total, there are twelve regions in the kinematic plane that 
are relevant to DIS $(2+1)j$-production. It turns out that the analytic 
continuation to all these regions can be obtained by deriving continuation 
formulae to four regions (which we take to be (1d)$_-$, (4d)$_-$, (4b)$_-$ and
(3c)$_-$), while the continuation to the remaining eight regions is then 
obtained using crossings of arguments, as described in the previous section. We
establish the continuation formulae for all these cases in this section.

\subsection{Continuation from (1a)$_-$ to (1d)$_-$ and to (1b)$_-$, 
(1c)$_-\ $} 

In region (1d)$_-$, 
which contributes to DIS in the momentum arrangement
$p_4+p_1 \to p_2 + p_3$ (but does not cover the full phase space available 
for this reaction), we have $ 0<y<1,\ \ 0 > z > -y $, \ie 
$\ \ -1 < z < 0,\ \ -z < y < 1 $, and 
\begin{equation}
q^2 <0, \; \quad s_{12} < 0,\; \qquad q^2 < s_{13}  < 0,\; \quad 
- s_{13} > s_{23}  >0 \; ,
\label{eq:1dkin}
\end{equation}
such that only $s_{23}$ needs to be assigned an infinitesimal imaginary part 
$+\iep$.
Owing to (\ref{eq:1dkin}), one has
\begin{equation}
\frac{s_{13}}{ s_{12}+s_{13}+s_{23}+\iep} = y+\iep\; ,\qquad 
\frac{s_{23}+\iep}{ s_{12}+s_{13}+s_{23}+\iep} = z-\iep\;.
\end{equation}
Equation~(\ref{eq:start}) therefore reads 
\begin{eqnarray}
\Psi(s_{12},s_{13},s_{23}+\iep) &=& (-q^2)^\alpha \Phi\left( 
     \frac{s_{13}}{ s_{12}+s_{13}+s_{23}+\iep}, 
     \frac{s_{23}+\iep}{ s_{12}+s_{13}+s_{23}+\iep} \right) \nonumber\\ 
    &=& (-q^2)^\alpha \Phi\left(y+\iep,z-\iep \right) \ . 
\end{eqnarray}

Note that the sign of the imaginary part associated with $y$, as will be 
clear from the following, actually plays no role 
in the assignment of imaginary parts to the 2dHPLs, since $y$ remains in the 
range $0<y<1<1-z$ in (1d)$_-$, which is free from cuts.

In region (1d)$_-$, we introduce the new variables
\begin{equation}
r_{{\rm 1d}} =r  =   \frac{s_{12}+s_{23}}{q^2} = 1-y , \qquad 
s_{{\rm 1d}} =s  =  - \frac{s_{23}}{q^2} = - z \; ,
\label{eq:1d+1}
\end{equation}
so that 
\begin{equation}
y = \frac{s_{13}}{ q^2} 
= 1-r  \;, 
\quad z =  \frac{ s_{23} }{ s_{12}+s_{13}+s_{23} } = -s\; ; 
\label{eq:start1d}
\end{equation} 
in the region (1d)$_-$, $(r,s)$ fulfil $0\leq r \leq 1$, $0\leq s \leq 1-r$, 
or $0\leq s \leq 1$, $0\leq r\leq 1-s$, thus mapping 
(1d)$_-$ onto (1a)$_-$. 

The generic expression for a 
master integral, Eq.~(\ref{eq:start}), then reads 
 \begin{equation}
\Psi(s_{12},s_{13},s_{23}+\iep) = 
      (-q^2)^\alpha \Phi\left( y + \iep , z - \iep\right) = 
      (-q^2)^\alpha \Phi\left( 1-r + \iep , -s - \iep\right) . 
\end{equation}

In terms of these variables, the HPLs of $w=1$ are continued as 
\begin{eqnarray}
\H(0;z-\iep) & = & \log\left(-s - \iep\right)
 =  \log\left(s \right) - i\pi \nonumber\\ 
& = & \H(0;s) - i\pi \nonumber \\ 
\H(1;z-\iep) & = & - \log\left(1+s+\iep\right)
             \nonumber \\ 
& = & - \H(-1;s)  \; ,
\label{eq:hconb}
\end{eqnarray}
and the 2dHPLs at $w=1$ as
\begin{eqnarray}
\G(0;y+\iep) & = & \log\left(1-r+\iep\right) \nonumber \\ 
& = & \G(1;r)  \nonumber \\
\G(1;y+\iep) & = & \log\left(r-\iep\right) \nonumber \\
& = & \G(0;r) \nonumber \\ 
\G(1-z+\iep;y+\iep) & = & \log\left(\frac{r-\iep+s+\iep}{1+s+\iep}\right) 
= \log\left(\frac{r+s}{s}\right) + \log(s) - \log(1+s) \nonumber \\
& = & \G(-s;r) + \H(0;s) - \H(-1;s) \nonumber\\
 \G(-z+\iep;y+\iep) 
& = & \log\left( 1 - \frac{1-r+\iep}{s+\iep} \right) 
    = \log\left( 1 - \frac{1-r}{s} \right) + i\pi \nonumber \\ 
& = & \G(1-s;r) - \H(1;s) - \H(0;s) + i\pi \; , 
\label{eq:gconb} 
\end{eqnarray} 
where we have used 
$$ \frac{1-r+\iep}{s+\iep} = \frac{1-r}{s} - \iep \ , $$ 
as $0\leq s \leq 1-r $ in the region (1d)$_-$. 
Using the above formulae, all imaginary parts of the higher weight 
HPL and 2dHPL are fixed, since these functions can be derived iteratively 
by integrating the $w=1$ functions. 

The HPLs with $w>1$ are obtained by first 
separating all rightmost $(0)$'s in the index vector using the product 
algebra~\cite{hpl}. The continuation to $ z - \iep = - s - \iep $ (flipping 
the sign of the argument) is then carried out (when the rightmost 
index $a_1$ is different from $0$) as described in~\cite{1dhpl}: 
\begin{eqnarray}
  \H(a_2,a_1;z-\iep) &=& (-1)^{a_1+a_2}  \H(-a_2,-a_1;s) 
                                         \ , \nonumber\\ 
  \H(a_3,a_2,a_1;z-\iep) &=& (-1)^{a_1+a_2+a_3} 
                   \H(-a_3,-a_2,-a_1;s) \ , \nonumber\\ 
  \H(a_4,a_3,a_2,a_1;z-\iep) &=& (-1)^{a_1+a_2+a_3+a_4} 
              \H(-a_4,-a_3,-a_2,-a_1;s) \ . 
\label{eq:xtoy2} 
\end{eqnarray} 
Notice that the $\iep$ in the argument is relevant only to the HPLs with 
rightmost $(0)$'s, which acquire an imaginary part in this transformation. 

To perform the continuation of the 2dHPLs with weight $w>1$, one can 
proceed by induction on the weight $w$, starting from the $w=1$ 
formulae (\ref{eq:gconb}). At $w>1$ one first separates off all leftmost 
$(1)$ components of the index vector by using the product 
algebra of the 2dHPLs, so that one has to consider only vector indices 
of the form $(a(z),\vec{b}(z))$, where $a(z)$ stands for one of the values 
$(0,1-z,-z)$, while $\vec{b}(z)$ can contain the index $(1)$ 
as well. By using the very definition of the 2dHPLs, one can write in 
full generality 
\begin{eqnarray} 
  \G(a(z),\vec{b}(z);y) &=& \int_0^{y} dy' \g(a(z);y') \G(\vec{b}(z);y') 
                                           \nonumber\\ 
  &=& \G(a(z),\vec{b}(z);1) + \int_1^{y} dy' \g(a(z);y') \G(\vec{b}(z);y') 
                                                                \nonumber\\ 
  &=& \G(a(z),\vec{b}(z);1) - \int_0^{1-y} dr' \g(a(z);1-r') 
                                                     \G(\vec{b}(z);1-r')\ , 
\label{eq:ggenb} 
\end{eqnarray} 
where the new integration variable $r'=1-y'$ has been introduced. Let 
us recall that $\G(a(z),\vec{b}(z);1) $ is finite as the leftmost index 
$a(z)$ is different from 1. 

The above formula is well suited for the continuation to 
$(z=-s-\iep,\ y=1-r+\iep)$: 
\begin{eqnarray} 
\lefteqn{\G(a(-s-\iep),\vec{b}(-s-\iep);1-r+\iep) = }\nonumber\\ 
 &=& \G(a(-s-\iep),\vec{b}(-s-\iep);1) - \int_0^{r-\iep} dr' 
     \g(a(-s-\iep);1-r') 
                         \G(\vec{b}(-s-\iep);1-r')\ \nonumber\\ 
 &=& \G(a(-s-\iep),\vec{b}(-s-\iep);1) - \int_0^{r} dr' 
          \g(a(-s-\iep);1-r'+\iep) \G(\vec{b}(-s-\iep);1-r'+\iep) \ . 
\label{eq:ggenbb} 
\end{eqnarray} 
The values at $y=1$ can be evaluated by expressing $\G(a(z),\vec{b}(z);1) $, 
for $0\le z\le 1$, in terms of HPLs of argument $z$, and then continuing 
the resulting expression to $z-\iep=-s-\iep$ as already discussed above, 
Eqs.~(\ref{eq:xtoy2}). 
For evaluating the integral, note that the integration variable runs in the 
region $0\le r' \le r \le 1-s$, so that 
the expressions of the rational fractions $\g(a(-s-\iep);1-r'+\iep) $ are 
\begin{eqnarray} 
\g(0;1-r'+\iep) & = &  - \g(1;r')\; , \nonumber \\ 
\g(1;1-r'+\iep) & = &  - \g(0;r')\; , \nonumber \\ 
\g(1+s+\iep;1-r'+\iep) & = & - \g(-s;r')\;, \nonumber\\ 
\g(s+\iep;1-r'+\iep) & = & -\g(1-s;r') \; , \nonumber 
\end{eqnarray} 
where the $\iep$ can be dropped as they are irrelevant in the 
considered region of $r'$. 
Finally, the last term appearing in (\ref{eq:ggenbb}), 
$\G(\vec{b}(-s-\iep);1-r'+\iep)$ is of weight $w-1$, so that its expression 
in terms of 2dHPLs depending on $(s,r')$ is also known on that region, 
and the $r'$-integral in (\ref{eq:ggenbb}) can be immediately evaluated 
in terms of 2dHPLs of argument $r$. 

An example for the continuation of a 2dHPL is
\begin{eqnarray}
\G(-z+\iep,1-z+\iep;y+\iep) &=& \G(1-s,-s;r) 
     + \left[ \H(0;s) - \H(-1;s)\right] \G(1-s;r) + \H(-1,0;s) \nonumber \\ 
&&          + \H(-1,1;s) 
     + \H(1,-1;s)
     - \H(1,0;s)
     - \frac{\pi^2}{6}
     - i\pi\; \H(-1;s) \ , 
\end{eqnarray} 
where use has been made of 
\begin{equation}
\G(-z+\iep,1-z+\iep;1) = \H(-1,0;s) + \H(-1,1;s) + \H(1,-1;s) 
                  - \H(1,0;s) - \frac{\pi^2}{6} - i\pi\; \H(-1;s) \ . 
\end{equation}
 
This example also illustrates the major new feature encountered in the 
continuation from (1a)$_-$ to (1d)$_-$ (or to any of the regions labelled by 
(b,c,d) in Fig.~\ref{fig:master_yz}): the cut structure of the boundaries 
of the (b,c,d)-type regions does not reproduce the cut structure of 
the boundaries of the (a)-type regions. At each corner of any (a)-type 
region, three cuts intersect, while all (b,c,d)-type regions 
have one corner with only two cuts intersecting, plus other corners 
(if any) with three cuts intersecting. 
In the case of the region (1d)$_-$, it is the lower 
right corner, which touches only two cuts ($y=1$ and $y=-z$), while the 
two upper corners touch three cuts each ($z=0$, $y=0$ and $y=-z$ for the 
upper left corner and $z=0$, $y=1$ and $y=1-z$ for the upper right corner 
respectively). Moreover, the (a)-type regions touch {\it all} cuts 
present in the HPLs and 2dHPLs on at least one of their corners,
which is not the case for the (b,c,d)-type regions. 
For this reason,
it is not possible to express the HPL and 2dHPL 
from (1a) in terms of HPL and 2dHPL with same set of elements in  
the index vector in the (b,c,d)-type region. 
The case just considered of the continuation 
from (1a)$_-$ to (1d)$_-$ is 
in this respect very fortunate, since the `missing' cut in $z=1$
(which is not touched by any boundary of (1d)$_-$) 
translates into a new cut in $s=-1$, which requires only an 
extension of the set of indices of the HPL from 
(0,1) to $(-1,0,1)$, while leaving the index set of the 2dHPL 
unaltered. The set $(-1,0,1)$ coincides with the original definition of the 
HPL~\cite{hpl}, and the numerical implementation~\cite{1dhpl} is 
covering this set.

It is clear that identities for the exchange and redefinition of arguments 
of the 2dHPL, which mix HPL and 2dHPL in a given region, owing to the 
presence of the value $(-1)$ in the vector of the indices of the 
HPLs, are no longer applicable in (1d)$_-$. 
An important consequence of this is 
furthermore that only a choice of variables of the form (\ref{eq:1d+1})
allows all functions in (1d)$_-$ to be expressed in terms of HPLs and 2dHPLs 
without an extension of the rational factors and the indices of the 
considered 2dHPLs. 
Hence, continuation to (1b,c)$_-$ cannot be performed in two ways (as in the
previous section), but only by 
using the second method, \ie~ redefining the independent variables 
in the Euclidean region. 

In region (1b)$_-$, ($0<z<1,\ 1-z<y<1$), we have
\begin{equation}
q^2 <0, \; \quad s_{12} < 0,\; \qquad q^2 < s_{23}  < 0,\; \quad 
- s_{23} > s_{13}  >0 \; ,
\label{eq:1bkin}
\end{equation}
so that imaginary parts have to be assigned to $s_{13}$ only. 

Continuation from (1a)$_-$ to (1b)$_-$ is performed by first re-expressing 
the HPLs $\H(\vec{m};z)$ and 2dHPLs $\G(\vec{m}(z);y)$ in (1a)$_-$ 
by HPLs $\H(\vec{m};x)$ and 2dHPLs $\G(\vec{m}(x);y)$, 
$ x = 1-y-z $, which is made according to (\ref{eq:startyx}).
The algorithm described 
in this section is then used to obtain the continuation to (1b)$_-$ in 
terms of HPLs $\H(\vec{m};s_{{\rm 1b}})$ and  2dHPLs 
$\G(\vec{m}(s_{{\rm 1b}});r_{{\rm 1b}})$, with 
\begin{equation}
r_{{\rm 1b}} =   \frac{s_{12}+s_{23}}{q^2} = 1-y , \qquad 
s_{{\rm 1b}} =  - \frac{s_{12}}{q^2} = - x \; ,
\end{equation}
which fulfil 
$0\leq s_{{\rm 1b}} \leq 1$, $0\leq r_{{\rm 1b}}\leq 1-s_{{\rm 1b}}$ in 
(1b)$_-$. For the reasons stated above, it is not possible to find a set of 
variables in (1b)$_-$, which makes the $(y\leftrightarrow z)$ symmetry in 
this region explicit and still retains the same set of indices for
HPLs and 2dHPLs.

Finally, in region (1c)$_-$, ($0<z<1,\ -z<y<0$), we have
\begin{equation}
q^2 <0, \; \quad s_{13} < 0,\; \qquad q^2 < s_{23}  < 0,\; \quad 
- s_{13} > s_{12}  >0 \; ,
\label{eq:1ckin}
\end{equation}
so that imaginary parts have to be assigned to $s_{12}$ only. 

Continuation from (1a)$_-$ to (1c)$_-$ follows similar lines by re-expressing 
the HPLs $\H(\vec{m};z)$ and 2dHPLs $\G(\vec{m}(z);y)$ in (1a)$_-$ 
by HPLs $\H(\vec{m};y)$ and 2dHPLs $\G(\vec{m}(y);z)$, using 
(\ref{eq:startzy}). The algorithm described 
in this section is then used to obtain the continuation to (1c)$_-$ in 
terms of HPLs $\H(\vec{m};s_{{\rm 1c}})$ and  2dHPLs 
$\G(\vec{m}(s_{{\rm 1c}});r_{{\rm 1c}})$, with 
\begin{equation}
r_{{\rm 1c}} = \frac{s_{12}+s_{13}}{q^2} = 1-z , \qquad 
s_{{\rm 1c}} = - \frac{s_{13}}{q^2} = - y \; ,
\end{equation}
which fulfil 
$0\leq s_{{\rm 1c}} \leq 1$, $0\leq r_{{\rm 1c}}\leq 1-s_{{\rm 1c}}$ 
in (1c)$_-$. 

\subsection{Continuation from (1a)$_-$ to (4d)$_-$ and to (2d)$_-$,(3d)$_-$ }

In region (4d)$_-$, $(1 < y < \infty, -\infty < z < -y)$, or 
$( -\infty < z < -1, 1 < y < -z)$, the invariants fulfil 
\begin{equation}
q^2 <0 ,\ \ s_{12} < q^2<0 ,\ \ s_{13} < q^2< 0 ,\ \ s_{23} > -q^2 > 0, 
\label{eq:4d+1}
\end{equation}
such that only $s_{23}$ is to be assigned an imaginary part $+\iep$.
Equation (\ref{eq:start}) reads as above in Section~\ref{sec:onecont}.1:
\begin{equation}
\Psi(s_{12},s_{13},s_{23}+\iep) = (-q^2)^\alpha \Phi\left( 
     \frac{s_{13}}{ s_{12}+s_{13}+s_{23}+\iep}, 
     \frac{s_{23}+\iep}{ s_{12}+s_{13}+s_{23}+\iep} \right) 
  = (-q^2)^\alpha \Phi(y+\iep,z-\iep) \ , 
\end{equation} 
as, because of Eqs.~(\ref{eq:4d+1}) and by definition of $y,z$ in 
Eqs.~(\ref{eq:defxyz}), 
\begin{equation}
 \frac{s_{13}}{s_{12}+s_{13}+s_{23}+\iep} 
= \frac{s_{13}}{q^2} +\iep = y +\iep \ ,\ \ \ 
 \frac{s_{23}+\iep}{s_{12}+s_{13}+s_{23}+\iep} = \frac{s_{23}}{q^2} -\iep 
= z -\iep \ . 
\label{eq:4d+3} 
\end{equation} 

To continue from (1a)$_-$ to (4d)$_-$, we introduce new dimensionless
variables, $r_{{\rm 4d}}$ and $s_{{\rm 4d}}$: 
\begin{equation}
r_{{\rm 4d}} =   \frac{s_{13}+s_{23}}{s_{23}} = \frac{y+z}{z} , \qquad 
s_{{\rm 4d}} = - \frac{q^2}{s_{23}} = - \frac{1}{z} \; ,
\label{eq:4d+4}
\end{equation}
and we express $y$ and $z$ in region (4d)
in terms of $r_{{\rm 4d}}$ and $s_{{\rm 4d}}$:
\begin{equation}
 y =  \frac{1-r_{{\rm 4d}}}{s_{{\rm 4d}}}  \ ,\ \ \ 
   z =  -\frac{1}{s_{{\rm 4d}}}  \ . 
\label{eq:yztors4d} 
\end{equation} 
As in all cases discussed before, $r_{{\rm 4d}}, s_{{\rm 4d}}$
 vary in the ranges 
$ 0 < s_{{\rm 4d}} < 1 $ and $ 0<r_{{\rm 4d}}<1-s_{{\rm 4d}}$, \ie\ the 
above parametrization maps region (4d)$_-$ into region (1a)$_-$. 
In these variables, the generic expression for a 
master integral, Eq.~(\ref{eq:start}), becomes 
 \begin{equation}
\Psi(s_{12},s_{13},s_{23}+\iep) = (-q^2)^\alpha \Phi\left( 
         \frac{1-r_{{\rm 4d}}}{s_{{\rm 4d}}} +\iep , 
        -\frac{1}{s_{{\rm 4d}}} - \iep\right) . 
\end{equation}

HPLs and 2dHPLs at weight $w=1$ thus become in (4d):
\begin{eqnarray}
\H(0;z-\iep) & = & \log\left(-\frac{1}{s_{{\rm 4d}}} - \iep\right)
 =  -\log\left(s_{{\rm 4d}} \right) - i\pi \nonumber\\ 
& = & -\H(0;s_{{\rm 4d}}) - i\pi \nonumber \\ 
\H(1;z-\iep) & = & - \log\left(1+\frac{1}{s_{{\rm 4d}}}+\iep\right)
             \nonumber \\ 
& = &  - \H(-1;s_{{\rm 4d}}) + \H(0;s_{{\rm 4d}}) \; ,\nonumber \\
\G(0;y+\iep) & = & \log\left(\frac{1-r_{{\rm 4d}}}{s_{{\rm 4d}}}+\iep\right)
\nonumber \\ 
& = & \G(1;r_{{\rm 4d}})-\H(0;s_{{\rm 4d}})  \nonumber \\
\G(1;y+\iep) & = & \log\left(1-\frac{1-r_{{\rm 4d}}}{s_{{\rm 4d}}}-\iep\right) 
= \log\left(\frac{1-r_{{\rm 4d}}-s_{{\rm 4d}}}{s_{{\rm 4d}}}\right) - i\pi\nonumber \\
& = & \G(1-s_{{\rm 4d}};r_{{\rm 4d}}) - \H(1;s_{{\rm 4d}})
-\H(0;s_{{\rm 4d}}) - i\pi \nonumber \\
\G(1-z+\iep;y+\iep) & = & \log\left(
\frac{1-\frac{1-r_{{\rm 4d}}}{s_{{\rm 4d}}}-\iep + \frac{1}{s_{{\rm 4d}}}+\iep}{1+\frac{1}{s_{{\rm 4d}}}+\iep}\right) 
\nonumber \\
& = & \G(-s_{{\rm 4d}};r_{{\rm 4d}}) + \H(0;s_{{\rm 4d}}) - \H(-1;s_{{\rm 4d}}) \nonumber\\
\G(-z+\iep;y+\iep) & = & \log\left(\frac{\frac{1-r_{{\rm 4d}}}{s_{{\rm 4d}}}+\iep-\frac{1}{s_{{\rm 4d}}}-\iep}
{-\frac{1}{s_{{\rm 4d}}}-\iep}\right)
\nonumber \\
& = & \G(0;r_{{\rm 4d}})  \; .
\label{eq:gcontesta}
\end{eqnarray}
The absence of imaginary parts from the last formula is non-trivial, and can 
be shown by explicitly inserting the definitions (\ref{eq:4d+4}), taking 
account of the boundaries (\ref{eq:4d+1}) on the invariants.

Continuation of the HPLs is made by combining the transformation 
formulae of Sections 5.1 (negation of argument) and 5.2 
(inversion of argument) of\cite{1dhpl}. 
Continuation of the 2dHPLs
requires first the separation of all leftmost $(-z)$'s in the index vector. 
The remaining 2dHPLs are then continued according to 
\begin{eqnarray}
&& {\kern-20pt} \G(a(z-\iep),\vec{b}(z-\iep);y+\iep) = \nonumber\\ 
&=& \G(a(z-\iep),\vec{b}(z-\iep);-z+\iep)
+ \int_{-z+\iep}^{y+\iep} \d y' \g(a(z-\iep),y') \G(\vec{b}(z-\iep);y') 
                                                            \nonumber\\ 
&=& \G(a(z-\iep),\vec{b}(z-\iep);-z+\iep) 
+ \int_{-z}^{y} \d y' \g(a(z-\iep),y'+\iep) \G(\vec{b}(z-\iep);y'+\iep) \ . 
\label{eq:ggenbx} 
\end{eqnarray} 
The term  $ \G(a(z-\iep),\vec{b}(z-\iep);-z+\iep) $  can be obtained by 
first evaluating $ \G(a(z),\vec{b}(z);-z) $, for $ 0 < z < 1 $, 
according to the algorithm described in the Appendix of~\cite{doublebox} 
and then continuing that result to $ z = - 1/s_{{\rm 4d}} -\iep$, yielding 
an expression containing HPL $\H(\vec{m};s_{{\rm 4d}})$. 
For the second term in Eq.~(\ref{eq:ggenbx}) one can introduce 
the integration variable $r'=1-s_{{\rm 4d}}y'$, 
or $y'=(1-r')/s_{{\rm 4d}}$, so that expressing 
$y,z$ in terms of $r_{{\rm 4d}},s_{{\rm 4d}}$, Eqs.~(\ref{eq:yztors4d}), the 
above equation becomes 
\begin{eqnarray} 
&& {\kern-20pt} \G(a(z-\iep),\vec{b}(z-\iep);y+\iep) = 
     \G\left(a\left( -\frac{1}{s_{{\rm 4d}}} -\iep\right), 
       \vec{b}\left( -\frac{1}{s_{{\rm 4d}}} -\iep\right); 
                     -\frac{1}{s_{{\rm 4d}}} +\iep\right) \nonumber\\ 
&& - \int_0^{r_{{\rm 4d}}} dr' \frac{1}{s_{{\rm 4d}}} 
 \g\left(a\left(-\frac{1}{s_{{\rm 4d}}}-\iep\right), \frac{1-r'}{s_{{\rm 4d}}} \right) 
 \ \G\left( \vec{b}\left(-\frac{1}{s_{{\rm 4d}}}-\iep\right); 
                                      \frac{1-r'}{s_{{\rm 4d}}} \right) \;,
\end{eqnarray} 
where the expression for $\G(\vec{b}(z);(1-r_{{\rm 4d}}')/s_{{\rm 4d}})$ in 
terms of 
$\G(\vec{c}(s_{{\rm 4d}});r_{{\rm 4d}}')$ and $\H(\vec{m};s_{{\rm 4d}})$
is again of lower weight and thus known. The expressions for the 
rational fractions are
\begin{eqnarray}
\frac{1}{s_{{\rm 4d}}} \g(0;y') & = &  - \g(1;r_{{\rm 4d}}')\; , \nonumber \\
\frac{1}{s_{{\rm 4d}}} \g(1;y') & = &  - \g(1-s_{{\rm 4d}};r_{{\rm 4d}}')\; , 
                                                                 \nonumber \\
\frac{1}{s_{{\rm 4d}}} \g(1-z;y') & = & -\g(-s_{{\rm 4d}};r_{{\rm 4d}}')\;, 
                                                                 \nonumber \\
\frac{1}{s_{{\rm 4d}}} \g(-z;y') & = & - \g(0;r_{{\rm 4d}}') \; ,
\end{eqnarray}
such that the $r_{{\rm 4d}}'$-integral  in 
(\ref{eq:ggenbx}) yields a 2dHPL of argument $r_{{\rm 4d}}$.

In region (2d)$_-$, $(1<y<\infty, 1<z<\infty)$, we have
\begin{equation}
q^2 <0 ,\ \ s_{13} < q^2<0 ,\ \ s_{23} < q^2< 0 ,\ \ s_{12} > -q^2 > 0, 
\end{equation} 
so that imaginary parts have to be assigned to $s_{12}$ only. 
For the continuation from (1a)$_-$ to (2d)$_-$, we use again in (1a)$_-$ 
the $x = 1-y-z$ or $s_{12} \leftrightarrow s_{13} $ interchange of 
Eq.~(\ref{eq:startyx}), so that its continuation to (2d)$_-$ is given by 
\begin{equation}
  \Psi(s_{12}+\iep,s_{13},s_{23}) =  ( -q^2 )^\alpha \Phi''\left( 
     \frac{s_{13}}{ s_{12}+s_{13}+s_{23}+\iep}, 
     \frac{s_{12}+\iep}{ s_{12}+s_{13}+s_{23}+\iep} \right) 
  = ( -q^2 )^\alpha \Phi''(y+\iep,x-\iep) \ . 
\label{eq:master2d} 
\end{equation} 
The algorithm already described 
in this section is then used to express $ \Phi''(y+\iep,x-\iep) $ in 
the region (2d)$_-$ in terms of HPLs $\H(\vec{m};s_{{\rm 2d}})$ and 
2dHPLs $\G(\vec{m}(s_{{\rm 2d}});r_{{\rm 2d}})$, with 
\begin{equation}
r_{{\rm 2d}} =   \frac{s_{12}+s_{13}}{s_{12}} = \frac{x+y}{x} , \qquad 
s_{{\rm 2d}} =  - \frac{q^2}{s_{12}} = - \frac{1}{x} \; ,
\end{equation}
which fulfil 
$0\leq s_{{\rm 2d}} \leq 1$, $0\leq r_{{\rm 2d}}\leq 1-s_{{\rm 2d}}$ in (2d).

Finally, in region (3d)$_-$, $( -\infty < y < -1,\ 1<z<-y) $ or 
$( 1<z<\infty,\ -\infty<y<-z)$, the invariants are bound by
\begin{equation}
q^2 <0 ,\ \ s_{12} < q^2<0 ,\ \ s_{23} < q^2< 0 ,\ \ s_{13} > -q^2 > 0, 
\end{equation}
so that imaginary parts have to be assigned to $s_{13}$ only. 
For the continuation from (1a)$_-$ to (3d)$_-$, we use again in (1a)$_-$ 
the $y \leftrightarrow z$ or $s_{13} \leftrightarrow s_{23} $ interchange of 
Eq.~(\ref{eq:startzy}), so that its continuation to (3d)$_-$ is given by 
\begin{equation}
  \Psi(s_{12},s_{13},s_{23}+\iep) =  ( -q^2 )^\alpha \Phi'\left( 
     \frac{s_{23}}{ s_{12}+s_{13}+s_{23}+\iep}, 
     \frac{s_{13}+\iep}{ s_{12}+s_{13}+s_{23}+\iep} \right) 
  = ( -q^2 )^\alpha \Phi'(z+\iep,y-\iep) \ . 
\label{eq:master3d} 
\end{equation} 

The algorithm described 
in this section is then used to express $ \Phi'(z+\iep,y-\iep) $ in the 
region (3d)$_-$ in terms of HPLs $\H(\vec{m};s_{{\rm 3d}})$ and  2dHPLs 
$\G(\vec{m}(s_{{\rm 3d}});r_{{\rm 3d}})$, with 
\begin{equation}
r_{{\rm 3d}} =\frac{s_{13}+s_{23}}{s_{13}} = \frac{y+z}{y} , \qquad 
s_{{\rm 3d}} = - \frac{q^2}{s_{13}} = - \frac{1}{y} \; ,
\end{equation}
which fulfil 
$0\leq s_{{\rm 3d}} \leq 1$, $0\leq r_{{\rm 3d}}\leq 1-s_{{\rm 3d}}$ in (3d).

\subsection{Continuation from (1a)$_-$ to (4b)$_-$ and to (2b)$_-$, (3b)$_-$} 

In region (4b)$_-$, $(1<z<\infty , 0<y+z<1)$, the invariants fulfil 
\begin{equation}
q^2 <0 ,\ \ s_{12} <0 ,\ \ s_{23} < q^2 ,\ \ 0 < s_{13} < -s_{23}\ , 
\label{eq:4b+1}
\end{equation}
such that only $s_{13}$ is to be assigned an imaginary part $+\iep$.
Equation~(\ref{eq:start}) then reads 
\begin{equation}
\Psi(s_{12},s_{13}+\iep,s_{23}) = (-q^2)^\alpha \Phi\left( 
     \frac{s_{13}+\iep}{ s_{12}+s_{13}+s_{23}+\iep}, 
     \frac{s_{23}}{ s_{12}+s_{13}+s_{23}+\iep} \right) 
     = (-q^2)^\alpha \Phi( y-\iep, z+\iep ) \ . 
\end{equation} 

For the continuation from (1a)$_-$ to (4b)$_-$ we introduce 
the dimensionless variables $r_{{\rm 4b}}$ and $s_{{\rm 4b}}$: 
\begin{equation}
r_{{\rm 4b}} =   \frac{s_{13}+s_{23}}{s_{23}} 
             = \frac{y+z}{z} \ , \qquad 
s_{{\rm 4b}} =   \frac{s_{12}}{s_{23}} = \frac{1-y-z}{z} \; ,
\label{eq:4b+3}
\end{equation}
such that
\begin{equation}
 y =  - \frac{1-r_{{\rm 4b}}}{r_{{\rm 4b}}+s_{{\rm 4b}}}  \ ,\ \ \ 
   z =  \frac{1}{r_{{\rm 4b}}+s_{{\rm 4b}}}  \ . 
\label{eq:4b+4}
\end{equation} 
As above
$ 0 < s_{{\rm 4b}} < 1 $ and $ 0<r_{{\rm 4b}}<1-s_{{\rm 4b}}$, 
thus mapping region (4d) into region (1a). 
In these variables, the generic expression for a 
master integral, Eq.~(\ref{eq:start}), becomes 
 \begin{equation}
\Psi(s_{12},s_{13}+\iep,s_{23}) = (-q^2)^\alpha \Phi\left( 
       - \frac{1-r_{{\rm 4b}}}{r_{{\rm 4b}}+s_{{\rm 4b}}} -\iep , 
       \frac{1}{r_{{\rm 4b}}+s_{{\rm 4b}}}  + \iep\right) . 
\end{equation}

The continuation of HPLs and 2dHPLs at weight $w=1$ reads:
\begin{eqnarray}
\H(0;z+\iep) & = & \log\left( \frac{1}{r_{{\rm 4b}}+s_{{\rm 4b}}}+ \iep\right)
=  -\log\left( \frac{r_{{\rm 4b}}+s_{{\rm 4b}}}{s_{{\rm 4b}}} \right)
+ \log s_{{\rm 4b}} \nonumber \\
& = & - \G(-s_{{\rm 4b}};r_{{\rm 4b}}) - \H(0;s_{{\rm 4b}}) \nonumber \\
\H(1;z+\iep) & = &-\log\left(1-\frac{1}{r_{{\rm 4b}}+s_{{\rm 4b}}}-\iep\right)
= -\log\left(\frac{1-{r_{{\rm 4b}}+s_{{\rm 4b}}}}
{{r_{{\rm 4b}}+s_{{\rm 4b}}}}\right) + i\pi \nonumber \\
& = & -\G(1-s_{{\rm 4b}};r_{{\rm 4b}}) + \G(-s_{{\rm 4b}};r_{{\rm 4b}})
    + \H(0;s_{{\rm 4b}}) + \H(1;s_{{\rm 4b}}) +i\pi \nonumber \\
\G(0;y-\iep) & = & \log\left(-\frac{1-r_{{\rm 4b}}}{r_{{\rm 4b}}+s_{{\rm 4b}}}
-\iep\right) = \log(1-r_{{\rm 4b}}) 
-\log(r_{{\rm 4b}}+s_{{\rm 4b}}) - i\pi \nonumber \\
& = & \G(1;r_{{\rm 4b}}) - \G(-s_{{\rm 4b}};r_{{\rm 4b}}) 
- \H(0;s_{{\rm 4b}})
-i\pi  \nonumber \\ 
\G(1;y-\iep) & = & \log\left(1+\frac{1-r_{{\rm 4b}}}{r_{{\rm 4b}}+s_{{\rm 4b}}}
+\iep\right) = \log\left(\frac{1+s_{{\rm 4b}}}
{r_{{\rm 4b}}+s_{{\rm 4b}}}\right) \nonumber \\
& =& -\G(-s_{{\rm 4b}};r_{{\rm 4b}})-\H(0;s_{{\rm 4b}})+\H(-1;s_{{\rm 4b}})
\nonumber \\
\G(1-z-\iep;y-\iep) & = & \log\left(\frac{1+
\frac{1-r_{{\rm 4b}}}{r_{{\rm 4b}}+s_{{\rm 4b}}}+\iep
-\frac{1}{r_{{\rm 4b}}+s_{{\rm 4b}}}- \iep}
{1-\frac{1}{r_{{\rm 4b}}+s_{{\rm 4b}}}- \iep}\right) = \log\left(
\frac{s_{{\rm 4b}}}{1-r_{{\rm 4b}}-s_{{\rm 4b}}}\right) + i\pi \nonumber \\
 & = & -\G(1-s_{{\rm 4b}};r_{{\rm 4b}}) +\H(0;s_{{\rm 4b}})+\H(1;s_{{\rm 4b}}) 
+i\pi \nonumber \\
\G(-z-\iep;y-\iep) & = & \log\left(\frac{-
\frac{1-r_{{\rm 4b}}}{r_{{\rm 4b}}+s_{{\rm 4b}}}-\iep
+\frac{1}{r_{{\rm 4b}}+s_{{\rm 4b}}}+ \iep}
{\frac{1}{r_{{\rm 4b}}+s_{{\rm 4b}}}+ \iep}\right) = \log(r_{{\rm 4b}})
\nonumber \\
& = & \G(0;r_{{\rm 4b}}) \;.
\label{eq:HG4b} 
\end{eqnarray}

At variance with all cases discussed before, one observes here that 
the 2dHPLs $\G(\vec{b}(s_{{\rm 4b}});r_{{\rm 4b}})$ appear not only in 
the continuation of the  2dHPLs
$\G(\vec{b}(z);y)$, but also in the continuation of the HPLs
$\H(\vec{b};z)$. This feature is due to the fact that $r_{{\rm 4b}}$
appears in the expressions for both $y$ and $z$ (\ref{eq:4b+4}), while it 
appeared only in the expression for $y$ in all cases discussed previously.
As a consequence, the continuations of $\G(\vec{b}(z);y)$ and $\H(\vec{b};z)$
of weights $w>1$ are more intertwined than in the cases discussed in 
previous sections, and no simple formulae for them can be 
given. Instead, these continuations have to be carried out in an 
algorithmic procedure, which is explained below. 

Continuation of the HPLs is made using
\begin{eqnarray}
\H(a,\vec{b};z+\iep) &=& \int_0^{z+\iep}\d z'\ \f(a,z')\H(\vec{b};z') 
                                                                \nonumber\\ 
 &=& \int_0^{\frac{1}{s_{{\rm 4b}}}+\iep}\d z'\ \f(a,z')\H(\vec{b};z') 
  + \int_{\frac{1}{s_{{\rm 4b}}}+\iep}^{z+\iep}\d z'\ \f(a,z')\H(\vec{b};z') 
                                                      \label{eq:Hcont4b} \\ 
& = & \H\left(a,\vec{b};\frac{1}{s_{{\rm 4b}}}+\iep\right)
-\int_0^{r_{{\rm 4b}}}\frac{\d r_{{\rm 4b}}'}{(r_{{\rm 4b}}'+s_{{\rm 4b}})^2}
\f\left( a;\frac{1}{r_{{\rm 4b}}'+s_{{\rm 4b}}} +\iep \right) \, 
\H\left(\vec{b};\frac{1}{r_{{\rm 4b}}'+s_{{\rm 4b}}}+\iep\right)\; , 
                                                                \nonumber 
\end{eqnarray} 
where the new integration variable $r_{{\rm 4b}}'$ was introduced with 
the substitution $ z' = 1/({r_{{\rm 4b}}'+s_{{\rm 4b}}})+\iep$. 
One finds for the rational fractions:
\begin{eqnarray} 
 \frac{1}{ (r_{{\rm 4b}}'+s_{{\rm 4b}})^2 } 
 \f\left(0;\frac{1}{r_{{\rm 4b}}'+s_{{\rm 4b}}}\right)
 &=& \g(- s_{{\rm 4b}},r_{{\rm 4b}}') \nonumber \\
 \frac{1}{ (r_{{\rm 4b}}'+s_{{\rm 4b}})^2 } 
\f\left(1;\frac{1}{r_{{\rm 4b}}'+s_{{\rm 4b}}}\right)
&=& \g(1-s_{{\rm 4b}},r_{{\rm 4b}}') - \g(- s_{{\rm 4b}},r_{{\rm 4b}}') \ . 
\end{eqnarray} 
The HPLs $\H(\vec{b};1/(r_{{\rm 4b}}'+s_{{\rm 4b}})+\iep)$ of 
Eq.~(\ref{eq:Hcont4b}) are of lower weight, and therefore known when 
proceeding bottom up starting from weight $w=1$. They can be expressed as
a linear combination of  HPLs $\H(\vec{c};s_{{\rm 4b}})$ and 2dHPLs
$\G(\vec{d}(s_{{\rm 4b}}),r_{{\rm 4b}}')$. As a consequence, the above
integral yields 2dHPLs $\G(\vec{e}(s_{{\rm 4b}}),r_{{\rm 4b}})$. Finally,
the boundary term $\H(a,\vec{b};1/{s_{{\rm 4b}}}
+\iep)$ is evaluated using the inversion formula of
Section 5.3 of~\cite{1dhpl}, yielding HPLs $\H(\vec{c};{s_{{\rm 4b}}})$.

For obtaining the continuation of the 2dHPLs, write 
\begin{equation} 
  \G(\vec{c}(z+\iep);y-\iep) = 
     \G\left(\vec{c}\left(\frac{1}{r_{{\rm 4b}}+s_{{\rm 4b}}}+\iep\right); 
      -\frac{1-r_{{\rm 4b}}}{r_{{\rm 4b}}+s_{{\rm 4b}}}-\iep\right) \ , 
\label{eq:Gcont4b} 
\end{equation} 
and observe that the r.h.s., considered as a function of $r_{{\rm 4b}}$, 
is equal to its value at $r_{{\rm 4b}}= 0 $ plus the integral of its 
derivative from $0$ to $r_{{\rm 4b}}$, \ie 
\begin{eqnarray} 
  \G(\vec{c}(z+\iep);y-\iep) &=& 
    \G\left(\vec{c}\left(\frac{1}{s_{{\rm 4b}}}+\iep\right); 
                 - \frac{1}{s_{{\rm 4b}}} -\iep\right) \nonumber\\ 
  &+& \int_0^{r_{{\rm 4b}}} \d r_{{\rm 4b}}'\, \frac{\d}{\d r_{{\rm 4b}}'}
  \G\left(\vec{c}\left(\frac{1}{r_{{\rm 4b}}'+s_{{\rm 4b}}}+\iep\right);
  -\frac{1-r_{{\rm 4b}}'}{r_{{\rm 4b}}'+s_{{\rm 4b}}}-\iep\right) \ . 
\label{eq:Gcont4b1} 
\end{eqnarray} 

The $r_{{\rm 4b}}'$-derivative in the above formula acts both on the 
argument of the 2dHPL and on the index vector. Writing out the 2dHPL 
in its multiple integral representation, this derivative can be carried 
out, yielding at most the squares of inverse rational factors. Using 
partial fractioning and integration by parts, the result of this 
differentiation can be rewritten as a linear combination of integral
representations of 2dHPLs $\G(\vec{d}(s_{{\rm 4b}}),r_{{\rm 4b}})$; 
the algebraic simplifications occurring in working out the arguments of 
the factors $\g(a,r'_{{\rm 4b}})$, appearing in the 
$r'_{{\rm 4b}}$-derivatives are similar to those encountered in the 
{\it r.h.s.} of Eq.~(\ref{eq:HG4b}). 
The boundary term $\G(\vec{c}(1/s_{{\rm 4b}}+\iep);-s_{{\rm 4b}}-\iep)$ 
is obtained by first working out $\G(\vec{c}(z);-z)$ for $0<z<1$ by the 
standard procedures described in the appendix of~\cite{doublebox}, 
then continuing the result to $z=-1/ s_{{\rm 4b}}$; it is expressed in 
terms of HPLs $\H(\vec{d};s_{{\rm 4b}})$.

In region (2b)$_-$, ($0<z<1,\ -\infty<y<-z$), 
\begin{equation}
q^2 <0 ,\ \ s_{23} <0 ,\ \ s_{12} < q^2 ,\ \ 0 < s_{13} < -s_{12}\ , 
\end{equation}
so that imaginary parts have to be assigned to $s_{13}$ only. 
Continuation from (1a)$_-$ to (2b)$_-$ uses again the $x=1-y-z$ 
or $ s_{12} \leftrightarrow s_{13} $ interchange of Eq.~(\ref{eq:startyx}), 
as in the continuation from (1a)$_-$ to (2d)$_-$, Eq.~(\ref{eq:master2d}). 
Subsequently, the algorithm described 
in this section is then used to obtain the continuation to (2b)$_-$ in 
terms of HPLs $\H(\vec{m};s_{{\rm 2b}})$ and  2dHPLs 
$\G(\vec{m}(s_{{\rm 2b}});r_{{\rm 2b}})$, with 
\begin{equation}
r_{{\rm 2b}} = \frac{s_{12}+s_{13}}{s_{12}} = \frac{x+y}{x} , \qquad 
s_{{\rm 2b}} = \frac{s_{23}}{s_{12}} = \frac{1-x-y}{x} \; ,
\end{equation}
which fulfil 
$0\leq s_{{\rm 2b}} \leq 1$, $0\leq r_{{\rm 2b}}\leq 1-s_{{\rm 2b}}$ in (2b).

Finally, in (3b)$_-$, ($1<y<\infty,\ -y<z<-y+1$), the invariants are bound by
\begin{equation}
q^2 <0 ,\ \ s_{12} <0 ,\ \ s_{13} < q^2 ,\ \ 0 < s_{23} < -s_{13}\ , 
\end{equation}
so that imaginary parts have to be assigned to $s_{23}$ only. 
Continuation from (1a)$_-$ to (3b)$_-$ employs again the 
$y\leftrightarrow z$ interchange of (\ref{eq:startzy}), as in the 
continuation from (1a)$_-$ to (3d)$_-$, Eq.~(\ref{eq:master3d}). 
The algorithm described 
in this section is then used to obtain the continuation to (3b)$_-$ in 
terms of HPLs $\H(\vec{m};s_{{\rm 3b}})$ and  2dHPLs 
$\G(\vec{m}(s_{{\rm 3b}});r_{{\rm 3b}})$, with 
\begin{equation}
r_{{\rm 3b}} =\frac{s_{13}+s_{23}}{s_{13}} = \frac{y+z}{y}, \qquad 
s_{{\rm 3b}} =\frac{s_{12}}{s_{13}} = \frac{1-y-z}{y} \; ,
\end{equation}
which fulfil 
$0\leq s_{{\rm 3b}} \leq 1$, $0\leq r_{{\rm 3b}}\leq 1-s_{{\rm 3b}}$ in (3b).

\subsection{Continuation from (1a)$_-$ to (3c)$_-$ and to (4c)$_-$ and 
(2c)$_-$} 

The last regions required for the kinematics of DIS
are (3c)$_-$ and (4c)$_-$, (2c)$_-$, related to it by crossings. 
In region (3c)$_-$, ($1<y<\infty,\ 0<z<1$) the invariants are bound by
\begin{equation}
q^2 <0 ,\ \ s_{12} > 0 ,\ \ s_{13} < q^2 <0,\ \ q^2< s_{23} < 0\ , 
\label{eq:3c+1}
\end{equation}
such that only $s_{12}$ is to be assigned an imaginary part $+\iep$.
Eq.~(\ref{eq:start}) then reads 
\begin{equation}
\Psi(s_{12}+\iep,s_{13},s_{23}) = (-q^2)^\alpha \Phi\left( 
     \frac{s_{13}}{ s_{12}+s_{13}+s_{23}+\iep}, 
     \frac{s_{23}}{ s_{12}+s_{13}+s_{23}+\iep} \right) 
     = (-q^2)^\alpha \Phi(y +\iep,z +\iep) .
\end{equation}

For the continuation from (1a)$_-$ to (3c)$_-$ we introduce 
the dimensionless variables 
are  $r_{{\rm 3c}}$ and $s_{{\rm 3c}}$: 
\begin{equation}
r_{{\rm 3c}} =  - \frac{s_{12}+s_{23}}{s_{13}} = \frac{y-1}{y}, \qquad 
s_{{\rm 3c}} =   \frac{s_{23}}{s_{13}} = \frac{z}{y} \; ,
\label{eq:3c+3}
\end{equation}
such that
\begin{equation}
 y =   \frac{1}{1-r_{{\rm 3c}}} \ ,\ \ \ 
   z =  \frac{s_{{\rm 3c}}}{1-r_{{\rm 3c}}}  \ . 
\label{eq:3c+4}
\end{equation} 
As above,
$ 0 < s_{{\rm 3c}} < 1 $ and $ 0<r_{{\rm 3c}}<1-s_{{\rm 3c}}$, 
thus mapping region (3c) into region (1a). 
In these variables, the generic expression for a 
master integral, Eq.~(\ref{eq:start}), becomes 
 \begin{equation}
\Psi(s_{12}+\iep,s_{13},s_{23}) = (-q^2)^\alpha \Phi\left( 
     \frac{1}{1-r_{{\rm 3c}}}+ \iep , 
      \frac{s_{{\rm 3c}}}{1-r_{{\rm 3c}}} + \iep\right) . 
\end{equation}

The HPLs and 2dHPLs at weight $w=1$ are continued as:
\begin{eqnarray}
\H(0;z+\iep) & = & \log\left( \frac{s_{{\rm 3c}}}{1-r_{{\rm 3c}}} + \iep\right)
\nonumber \\
& = & \H(0;s_{{\rm 3c}}) - \G(1;r_{{\rm 3c}}) \nonumber \\
\H(1;z+\iep) & = & -\log\left( 1-\frac{s_{{\rm 3c}}}{1-r_{{\rm 3c}}} 
- \iep\right) \nonumber\\
& = & -\G(1-s_{{\rm 3c}};r_{{\rm 3c}}) + \G(1;r_{{\rm 3c}}) 
       + \H(1;s_{{\rm 3c}}) 
        \nonumber \\
\G(0;y+\iep) & = & \log\left( \frac{1}{1-r_{{\rm 3c}}} + \iep\right)
\nonumber \\
& = & -\G(1;r_{{\rm 3c}}) \nonumber \\
\G(1;y+\iep) & = & \log\left(1- \frac{1}{1-r_{{\rm 3c}}} - \iep\right)
= \log\left(\frac{r_{{\rm 3c}}}{1-r_{{\rm 3c}}}\right) - i\pi \nonumber \\
& = & \G(0;r_{{\rm 3c}}) -\G(1;r_{{\rm 3c}}) -i\pi \nonumber\\
\G(1-z-\iep;y+\iep) & = & \log\left(\frac{1-\frac{1}{1-r_{{\rm 3c}}} -\iep
-\frac{s_{{\rm 3c}}}{1-r_{{\rm 3c}}} -\iep}
{1-\frac{s_{{\rm 3c}}}{1-r_{{\rm 3c}}} -\iep}\right) 
= \log\left(\frac{r_{{\rm 3c}}+s_{{\rm 3c}}}{1-r_{{\rm 3c}}-s_{{\rm 3c}}}
\right) -i\pi \nonumber \\
& = & \G(-s_{{\rm 3c}};r_{{\rm 3c}}) - \G(1-s_{{\rm 3c}};r_{{\rm 3c}})
                  + \H(0;s_{{\rm 3c}}) + \H(1;s_{{\rm 3c}}) 
             - i\pi \ , \nonumber \\
\G(-z-\iep;y+\iep) & = & \log \left( \frac{\frac{1}{1-r_{{\rm 3c}}} +\iep
+\frac{s_{{\rm 3c}}}{1-r_{{\rm 3c}}} +\iep}
{\frac{s_{{\rm 3c}}}{1-r_{{\rm 3c}}} +\iep} \right)
= \log\left(\frac{1+s_{{\rm 3c}}}{ s_{{\rm 3c}}}\right) \nonumber \\
& = & \H(-1;s_{{\rm 3c}}) - \H(0;s_{{\rm 3c}}) \ . 
\end{eqnarray}
As in the continuation to (4b)$_-$, one observes here that 2dHPLs 
$\G(\vec{b}(s_{{\rm 3c}});r_{{\rm 3c}})$ appear both in the 
continuation of the  2dHPLs $\G(\vec{b}(z);y)$ and  of the HPLs 
$\H(\vec{b};z)$. The continuation of the higher-weight functions also follows 
 similar lines as for (4b)$_-$.

As in Eq.~(\ref{eq:Hcont4b}), continuation of the HPLs is made using
\begin{eqnarray}
\H(a,\vec{b};z+\iep) & = & \H\left(a,\vec{b};s_{{\rm 3c}} +\iep\right)
  + \int_{s_{{\rm 3c}}+\iep}^{z+\iep} \d z' \f(a,z')\H(\vec{b};z') 
      \label{eq:Hcont3c} \\
                     & = & \H\left(a,\vec{b};s_{{\rm 3c}} +\iep\right) 
  +\int_0^{r_{{\rm 3c}}}\d r_{{\rm 3c}}' 
       \frac{s_{{\rm 3c}}}{(1-r_{{\rm 3c}}')^2} 
       \f\left(a;\frac{s_{{\rm 3c}}}{1-r_{{\rm 3c}}'}+\iep\right) \, 
 \H\left(\vec{b};\frac{s_{{\rm 3c}}}{1-r_{{\rm 3c}}'}+\iep\right)\; . 
                                                     \nonumber 
\end{eqnarray}
One finds for the rational fractions:
\begin{eqnarray}
    \frac{s_{{\rm 3c}}}{(1-r_{{\rm 3c}}')^2} 
    \f\left(0;\frac{s_{{\rm 3c}}}{1-r_{{\rm 3c}}'} +\iep \right)
   &=& - \g(1,r_{{\rm 3c}}') \ , \nonumber \\
    \frac{s_{{\rm 3c}}}{(1-r_{{\rm 3c}}')^2} 
    \f\left(1;\frac{s_{{\rm 3c}}}{1-r_{{\rm 3c}}'} +\iep \right)
   &=& \g(1;r_{{\rm 3c}}') - \g(1-s_{{\rm 3c}};r_{{\rm 3c}}') \ . 
\end{eqnarray}
The HPLs $\H(\vec{b};s_{{\rm 3c}}/(1-r_{{\rm 3c}}')+\iep)$ are 
of lower weight, and therefore known. They can be expressed as
linear combination of  HPLs $\H(\vec{c};s_{{\rm 3c}})$ and 2dHPLs
$\G(\vec{d}(s_{{\rm 3c}}),r_{{\rm 3c}}')$. As a consequence, the above
integral yields 2dHPLs $\G(\vec{e}(s_{{\rm 3c}}),r_{{\rm 3c}})$. 

To continue the  2dHPLs, following Eqs.~(\ref{eq:Gcont4b})
and (\ref{eq:Gcont4b1}) 
write 
\begin{equation} 
 \G(\vec{b}(z+\iep);y+\iep) =  \G\left( 
    \vec{b}\left(\frac{s_{{\rm 3c}}}{1-r_{{\rm 3c}}}+\iep\right);
    \frac{1}{1-r_{{\rm 3c}}}+\iep \right) 
\end{equation} 
and observe that the r.h.s., considered as a function of 
$r_{{\rm 3c}}$, is equal to its value at $r_{{\rm 3c}}=0$ plus the integral 
of its derivative from $0$ to $r_{{\rm 3c}}$, \ie 
\begin{eqnarray} 
\G(\vec{b}(z+\iep);y+\iep) & = & 
   \G\left(\vec{b}\left(s_{{\rm 3c}}+\iep\right);1+\iep\right) \nonumber \\ 
  && + \int_0^{r_{{\rm 3c}}} \d r_{{\rm 3c}}'
     \, \frac{\d}{\d r_{{\rm 3c}}'} 
    \G\left(\vec{b}\left(\frac{s_{{\rm 3c}}}{1-r_{{\rm 3c}}'}+\iep\right);
    \frac{1}{1-r_{{\rm 3c}}}+\iep\right) \ . 
\end{eqnarray} 

As in the previous section, the $r_{{\rm 3c}}'$-derivative 
in the above formula acts both on the argument of 
the 2dHPL and on the index vector. Writing out the 2dHPL 
in its multiple-integral representation, this derivative can be carried 
out, yielding at most the squares of inverse rational factors. Using 
partial fractioning and integration by parts, the result of this 
differentiation can be rewritten as a linear combination of integral
representations of 2dHPLs $\G(\vec{c}(s_{{\rm 3c}}),r_{{\rm 3c}})$.
The boundary term $\G(\vec{b}(s_{{\rm 3c}}+\iep);1+\iep)$ is obtained 
using standard procedures, as described in the appendix of~\cite{doublebox},
yielding $\H(\vec{c};s_{{\rm 3c}})$. 

In region (4c)$_-$, ($0<y<1,\ 1<z<\infty$), 
\begin{equation}
q^2 <0 ,\ \ s_{12} > 0 ,\ \ s_{23} < q^2 <0,\ \ q^2< s_{13} < 0\ , 
\end{equation}
so that imaginary parts have to be assigned to $s_{12}$ only. 
Continuation from (1a)$_-$ to (4c)$_-$ uses again the $z\leftrightarrow 
y$ interchange of (\ref{eq:master3a}). Subsequently, 
the algorithm described 
in this section is then used to obtain the continuation to (4c)$_-$ in 
terms of HPLs $\H(\vec{m};s_{{\rm 4c}})$ and  2dHPLs 
$\G(\vec{m}(s_{{\rm 4c}});r_{{\rm 4c}})$, with 
\begin{equation}
r_{{\rm 4c}} = - \frac{s_{12}+s_{13}}{s_{23}} = \frac{z-1}{z}\ , \qquad 
s_{{\rm 4c}} =  \frac{s_{13}}{s_{23}} = \frac{y}{z} \; ,
\end{equation}
which fulfil 
$0\leq s_{{\rm 4c}} \leq 1$, $0\leq r_{{\rm 4c}}\leq 1-s_{{\rm 4c}}$ in (4c).

Finally, in region (2c)$_-$, ($0<y<1,\ -\infty<z<-y$), the conditions 
on the invariants are 
\begin{equation}
q^2 <0 ,\ \ s_{23} >0 ,\ s_{12} < q^2 <0,\ \ q^2< s_{13} < 0\ , 
\end{equation}
so that imaginary parts have to be assigned to $s_{23}$ only. 
Continuation from (1a)$_-$ to (4c)$_-$ employs as well the 
$y\leftrightarrow z$ interchange of (\ref{eq:master3a}), followed by
the $z = 1-x-y $ replacement of (\ref{eq:master2a}), which maps 
(4c) into ($1<x<\infty,\ 0<y<1$); note that 
this procedure, which involves both the interchange and the replacement 
of the arguments, differs from all crossings discussed up to here.  
The algorithm described 
in this section is then used to obtain the continuation to (2c)$_-$ in 
terms of HPLs $\H(\vec{m};s_{{\rm 2c}})$ and 2dHPLs 
$\G(\vec{m}(s_{{\rm 2c}});r_{{\rm 2c}})$, with 
\begin{equation} 
  r_{{\rm 2c}} =  -\frac{s_{13}+s_{23}}{s_{12}} = \frac{x-1}{x} , \qquad 
  s_{{\rm 2c}} = \frac{s_{13}}{s_{12}} = \frac{y}{x} \; , 
\end{equation}
which fulfil 
$0\leq s_{{\rm 2c}} \leq 1$, $0\leq r_{{\rm 2c}}\leq 1-s_{{\rm 2c}}$ in (2c).

\section{Conclusions}
\setcounter{equation}{0}
\label{sec:conc}
In this paper, we have described the analytic continuation of two-loop 
four-point functions with massless internal propagators and one off-shell 
external leg from the Euclidean region to all Minkowskian regions of physical 
interest. While the continuation to the region relevant to $1\to 3$ kinematics
(as in $e^+e^- \to 3$~jets) amounts to simply continuing an overall 
factor~\cite{doublebox,3jme}, the continuations for the $2\to 2$ scattering 
processes ($V+1j$ and DIS-$(2+1)j$ production) is more involved. In particular,
since all two-loop four-point functions are expressed in terms of 2dHPLs 
whose arguments lie in general outside the analyticity range, for which 
numerical routines are available~\cite{2dhpl}, we had to find appropriate 
variable substitutions to map each kinematical region into the analyticity 
range of the 2dHPLs. These variable transformations are summarized 
in Table~\ref{tab:one}. 
\begin{table}[t]
{
\begin{center}
\vspace{0mm}
\begin{tabular}{|c|c|c|l|}\hline
\rule[0mm]{0mm}{3mm}
\raisebox{-1.5ex}[1.5ex]{Region} 
& \multicolumn{2}{|c|}{\rule[0mm]{0mm}{3mm}\raisebox{-1ex}[1ex]{ Variables}} 
&\multicolumn{1}{|c|}{\rule[0mm]{0mm}{3mm}\raisebox{-1.5ex}[1.5ex]{Procedure}}
\\
& $y$-type & $z$-type &   \\[1mm] \hline
\rule[-3mm]{0mm}{8mm}
(1a) & $y_{{\rm 1a}} = \frac{{\textstyle\sac}}{{\textstyle q^2}}$ & 
       $z_{{\rm 1a}} =\frac{{\textstyle\sbc}}{{\textstyle q^2}}$ &
  \\[1mm] \hline 
\rule[-3mm]{0mm}{8mm}
(1b) & $r_{{\rm 1b}} = \frac{{\textstyle \sab+\sbc}}{{\textstyle q^2}}$
  & $s_{{\rm 1b}}=- \frac{{\textstyle \sab}}{{\textstyle q^2}}$ 
                & ($y_{{\rm 1a}},z_{{\rm 1a}}$)
$\to$  ($y_{{\rm 1a}},1-y_{{\rm 1a}}-z_{{\rm 1a}}$)
$\to$  ($r_{{\rm 1b}},s_{{\rm 1b}}$) \\[1mm] \hline 
\rule[-3mm]{0mm}{8mm}
(1c) & $r_{{\rm 1c}} = \frac{{\textstyle \sab+\sac}}{{\textstyle q^2}}$
  & $s_{{\rm 1c}}=- \frac{{\textstyle \sac}}{{\textstyle q^2}}$  
                & ($y_{{\rm 1a}},z_{{\rm 1a}}$) 
$\to$  ($z_{{\rm 1a}},y_{{\rm 1a}}$) 
$\to$  ($r_{{\rm 1c}},s_{{\rm 1c}}$) \\[1mm] \hline 
\rule[-3mm]{0mm}{8mm}
(1d) & $r_{{\rm 1d}} = \frac{{\textstyle \sab+\sbc}}{{\textstyle q^2}}$ 
     & $s_{{\rm 1d}}=- \frac{{\textstyle \sbc}}{{\textstyle q^2}}$ 
                & ($y_{{\rm 1a}},z_{{\rm 1a}}$)
$\to$  ($r_{{\rm 1d}},s_{{\rm 1d}}$) \\[1mm] \hline 
\rule[-3mm]{0mm}{8mm}
(2a) & $u_{{\rm 2a}} =- \frac{{\textstyle \sac}}{{\textstyle \sab}}$
  & $v_{{\rm 2a}}=  \frac{{\textstyle q^2}}{{\textstyle \sab}}$ 
                & ($y_{{\rm 1a}},z_{{\rm 1a}}$)
$\to$  ($y_{{\rm 1a}},1-y_{{\rm 1a}}-z_{{\rm 1a}}$)  
$\to$  ($u_{{\rm 2a}},v_{{\rm 2a}}$) \\[1mm] \hline 
\rule[-3mm]{0mm}{8mm}
(2b) & $r_{{\rm 2b}} = \frac{{\textstyle \sab+\sac}}{{\textstyle \sab}}$
  & $s_{{\rm 2b}}=   \frac{{\textstyle \sbc}}{{\textstyle \sab}}$ & 
                 ($y_{{\rm 1a}},z_{{\rm 1a}}$) 
$\to$  ($y_{{\rm 1a}},1-y_{{\rm 1a}}-z_{{\rm 1a}}$)
$\to$   ($r_{{\rm 2b}},s_{{\rm 2b}}$) \\[1mm] \hline 
\rule[-3mm]{0mm}{8mm}
(2c) & $r_{{\rm 2c}} = - \frac{{\textstyle \sac+\sbc}}{{\textstyle \sab}}$
  & $s_{{\rm 2c}}=   \frac{{\textstyle \sac}}{{\textstyle \sab}}$ 
                & ($y_{{\rm 1a}},z_{{\rm 1a}}$)
$\to$  ($z_{{\rm 1a}},y_{{\rm 1a}}$) 
$\to$  ($z_{{\rm 1a}},1-y_{{\rm 1a}}-z_{{\rm 1a}}$) 
$\to$  ($r_{{\rm 2c}},s_{{\rm 2c}}$) \\[1mm] \hline 
\rule[-3mm]{0mm}{8mm}
(2d) & $r_{{\rm 2d}} = \frac{{\textstyle \sab+\sac}}{{\textstyle \sab}}$
  & $s_{{\rm 2d}}=  - \frac{{\textstyle q^2}}{{\textstyle \sab}}$ 
                & ($y_{{\rm 1a}},z_{{\rm 1a}}$)
$\to$  ($y_{{\rm 1a}},1-y_{{\rm 1a}}-z_{{\rm 1a}}$)
$\to$  ($r_{{\rm 2d}},s_{{\rm 2d}}$) \\[1mm] \hline 
\rule[-3mm]{0mm}{8mm}
(3a) & $u_{{\rm 3a}} = -\frac{{\textstyle \sbc}}{{\textstyle \sac}}$
  & $v_{{\rm 3a}}=  \frac{{\textstyle q^2}}{{\textstyle \sac}}$ 
                & ($y_{{\rm 1a}},z_{{\rm 1a}}$)
$\to$  ($z_{{\rm 1a}},y_{{\rm 1a}}$) 
$\to$  ($u_{{\rm 3a}},v_{{\rm 3a}}$) \\[1mm] \hline 
\rule[-3mm]{0mm}{8mm}
(3b) & $r_{{\rm 3b}} = \frac{{\textstyle \sac+\sbc}}{{\textstyle \sac}}$
  & $s_{{\rm 3b}}=  \frac{{\textstyle \sab}}{{\textstyle \sac}}$  
                & ($y_{{\rm 1a}},z_{{\rm 1a}}$)
$\to$  ($z_{{\rm 1a}},y_{{\rm 1a}}$) 
$\to$  ($r_{{\rm 3b}},s_{{\rm 3b}}$) \\[1mm] \hline 
\rule[-3mm]{0mm}{8mm}
(3c) & $r_{{\rm 3c}} = - \frac{{\textstyle \sab+\sbc}}{{\textstyle \sac}}$
  & $s_{{\rm 3c}}=  \frac{{\textstyle \sbc}}{{\textstyle \sac}}$ 
                & ($y_{{\rm 1a}},z_{{\rm 1a}}$)
$\to$  ($r_{{\rm 3c}},s_{{\rm 3c}}$) \\[1mm] \hline 
\rule[-3mm]{0mm}{8mm}
(3d) & $r_{{\rm 3d}} = \frac{{\textstyle \sac+\sbc}}{{\textstyle \sac}}$
  & $s_{{\rm 3d}}=  -\frac{{\textstyle q^2}}{{\textstyle \sac}}$ 
                & ($y_{{\rm 1a}},z_{{\rm 1a}}$)
$\to$  ($z_{{\rm 1a}},y_{{\rm 1a}}$) 
$\to$  ($r_{{\rm 3d}},s_{{\rm 3d}}$) \\[1mm] \hline 
\rule[-3mm]{0mm}{8mm}
(4a) & $u_{{\rm 4a}} = -\frac{{\textstyle \sac}}{{\textstyle \sbc}}$ 
     & $v_{{\rm 4a}}=  \frac{{\textstyle q^2}}{{\textstyle \sbc}}$ 
                & ($y_{{\rm 1a}},z_{{\rm 1a}}$)
$\to$  ($u_{{\rm 4a}},v_{{\rm 4a}}$) \\[1mm] \hline 
\rule[-3mm]{0mm}{8mm}
(4b) & $r_{{\rm 4b}} = \frac{{\textstyle \sac+\sbc}}{{\textstyle \sbc}}$ 
     & $s_{{\rm 4b}}=  \frac{{\textstyle \sab}}{{\textstyle \sbc}}$ 
                & ($y_{{\rm 1a}},z_{{\rm 1a}}$)
$\to$  ($r_{{\rm 4b}},s_{{\rm 4b}}$) \\[1mm] \hline 
\rule[-3mm]{0mm}{8mm}
(4c) & $r_{{\rm 4c}} = - \frac{{\textstyle \sab+\sac}}{{\textstyle \sbc}}$ 
     & $s_{{\rm 4c}}=  \frac{{\textstyle \sac}}{{\textstyle \sbc}}$ 
                & ($y_{{\rm 1a}},z_{{\rm 1a}}$)
$\to$  ($z_{{\rm 1a}},y_{{\rm 1a}}$) 
$\to$  ($r_{{\rm 4c}},s_{{\rm 4c}}$) \\[1mm] \hline 
\rule[-3mm]{0mm}{8mm}
(4d) & $r_{{\rm 4d}} = \frac{{\textstyle \sac+\sbc}}{{\textstyle \sbc}}$ 
     & $s_{{\rm 4d}}=- \frac{{\textstyle q^2}}{{\textstyle \sbc}}$  
                & ($y_{{\rm 1a}},z_{{\rm 1a}}$)
$\to$  ($r_{{\rm 4d}},s_{{\rm 4d}}$) \\[1mm] \hline 
\end{tabular}
\end{center}
\label{tab:one}
\caption{Variables used for 2dHPLs and HPLs  in 
each kinematic region, and variable transformations applied in the analytic 
continuation from (1a) to each region.}
}
\end{table}

Each continuation of the two-loop four-point functions is performed by 
first identifying the variables crossing  a kinematical cut. Subsequently,
the basis functions of the two-loop four-point functions (HPLs and 2dHPLs)
are continued by first continuing the weight $w=1$ functions (which are 
just logarithms with a known and well-defined continuation), 
the functions of higher weight are then constructed using the product 
algebra and the integral representations 
of the HPLs and 2dHPLs. It turns out, using this approach, that all imaginary 
parts arising in the analytic continuation are made explicit, and their 
signs are fixed from the continuation of the $w=1$ functions, which are 
all listed in the appropriate sections. 

Using the continuation formulae derived in this paper, one can use the 
results obtained for the two-loop QCD corrections to the 
$e^+e^- \to 3$~jets matrix element and helicity amplitudes~\cite{3jme}
to derive the corresponding corrections to the matrix elements for 
vector-boson-plus-jet production at hadron colliders and 
to deep inelastic two-plus-one-jet production~\cite{newme}. This work is 
currently in progress.

\begin{appendix}
\renewcommand{\theequation}{\mbox{\Alph{section}.\arabic{equation}}}
\section{Harmonic polylogarithms}
\setcounter{equation}{0}
\label{app:hpl}
The generalized polylogarithms ${\rm S}_{n,p}(x)$ 
of Nielsen~\cite{nielsen} turn out to be insufficient for the computation of 
multi-scale integrals beyond one loop. To overcome this limitation, 
one has to extend generalized polylogarithms to harmonic 
polylogarithms~\cite{hpl,doublebox}.

Harmonic polylogarithms are obtained by the repeated integration of 
rational factors. If these rational factors contain, besides the 
integration variable, only constants, the resulting 
functions are one-dimensional harmonic polylogarithms 
(or simply harmonic polylogarithms, HPLs)\cite{hpl,1dhpl}. 
If the rational 
factors depend on a further variable, one obtains 
two-dimensional harmonic polylogarithms (2dHPLs)~\cite{doublebox,2dhpl}.
In the following, we recall the definition of both classes of functions, 
and summarize their properties. 

\subsection{One-dimensional harmonic polylogarithms}

The HPLs, introduced in \cite{hpl}, are 
one-variable functions $ \H(\vec{a};x) $ depending, besides the argument 
$x$, on a set of indices, grouped for convenience into the vector 
$\vec{a}$, whose components can take one of the three values $(1,0,-1)$ 
and whose number is the weight $w$ of the HPL. More explicitly, the three 
HPLs  with $w=1$ are defined as 
\begin{eqnarray} 
  \H(1;x) &=& \int_0^x \frac{\d x'}{1-x'} = - \ln(1-x) \ , \nonumber\\ 
  \H(0;x) &=& \ln x \ ,          \nonumber\\ 
  \H(-1;x) &=& \int_0^x \frac{\d x'}{1+x'} = \ln(1+x) \ ; 
\label{eq:defineh1} 
\end{eqnarray} 
their derivatives can be written as 
\begin{equation} 
  \frac{\d }{\d x} \H(a;x) = \f(a;x) \ , \hskip 1cm a=1,0,-1 \;,
\label{eq:derive1} 
\end{equation} 
where the 3 rational fractions $\f(a;x)$ are given by 
\begin{eqnarray} 
   \f(1;x) &=& \frac{1}{1-x} \ , \nonumber\\ 
   \f(0;x) &=& \frac{1}{x} \ , \nonumber\\ 
   \f(-1;x) &=& \frac{1}{1+x} \ . 
\label{eq:definef} 
\end{eqnarray} 
For weight $w$ larger than 1, write $ \vec{a} = (a, \vec b) $, where 
$a$ is the leftmost component of  $ \vec{a} $ and $\vec b $ stands 
for the vector of the remaining $(w-1)$ components. The harmonic 
polylogarithms of weight $w$ are then defined as follows: 
if all the $w$ components of $\vec a$ take the value 0, $\vec a$ is said 
to take the value $\vec 0_w$ and 
\begin{equation} 
\H(\vec{0}_w;x) = \frac{1}{w!} \ln^w{x} \ , 
\label{eq:defh0} 
\end{equation} 
while, if $\vec{a} \neq \vec{0}_w$,
\begin{equation} 
\H(\vec{a};x) = \int_0^x \d x' \ \f(a;x') \ \H(\vec{b};x') \ . 
\label{eq:defn0} 
\end{equation} 
In any case the derivatives can be written in the compact form 
\begin{equation} 
\frac{\d }{\d x} \H(\vec{a};x) = \f(a;x) \H(\vec{b};x) \ , 
\label{eq:derive} 
\end{equation} 
where, again, $a$ is the leftmost component of $ \vec a $ and 
$ \vec b $ stands for the remaining $(w-1)$ components. 
\par 
It is immediate to see, from the very definition Eq.\ (\ref{eq:defn0}), that 
there are $3^w$ HPLs  of weight $w$, and that they are linearly 
independent. The HPLs  are generalizations of Nielsen's 
polylogarithms~\cite{nielsen}. The function $ \S_{n,p}(x) $, in 
Nielsen's notation, is equal to the HPL whose first $n$ indices are all 
equal to 0 and the remaining $p$ indices all equal to 1:
\begin{equation}
   \S_{n,p}(x) = \H(\vec{0}_{n},\vec{1}_p;x) \; ; 
\end{equation} 
in particular the Euler polylogarithms $ \Li_n(x) = \S_{n-1,1}(x) $ 
correspond to 
\begin{equation}
   \Li_n(x) = \H(\vec{0}_{n-1},1;x) \; .
\end{equation} 

As shown in \cite{hpl}, the product of two HPLs  of a same argument $x$ 
and weights $p, q$ can be expressed as a combination of HPLs  of that 
argument and weight $r=p+q$, according to the product identity 
\begin{eqnarray} 
 \H(\vec{p};x)\H(\vec{q};x) & = & 
  \sum_{\vec{r} = \vec{p}\uplus \vec{q}} \H(\vec{r};x) \; , 
\label{eq:halgebra} \end{eqnarray} 
where $\vec p, \vec q$ stand for the $p$ and $q$ components of the indices 
of the two HPLs, while $\vec{p}\uplus \vec{q}$ represents all mergers of 
$\vec{p}$ and $\vec{q}$ into the vector $\vec{r}$ with $r$ components, 
in which the relative orders of the elements of $\vec{p}$ and $\vec{q}$ 
are preserved. 
\par 
The explicit formulae relevant up to weight 4 are 
\begin{equation} 
   \H(a;x) \; \H(b;x) =  \H(a,b;x) + \H(b,a;x) \ , 
\label{eq:alg0} 
\end{equation} 
\begin{eqnarray} 
   \H(a;x) \; \H(b,c;x) &=&  \H(a,b,c;x) + \H(b,a,c;x) + \H(b,c,a;x) 
                             \ , \nonumber\\ 
   \H(a;x) \; \H(b,c,d;x) &=&  \H(a,b,c,d;x) + \H(b,a,c,d;x) + \H(b,c,a,d;x) 
                              + \H(b,c,d,a;x) \ , 
\label{eq:alg1} 
\end{eqnarray} 
and 
\begin{eqnarray} 
  \H(a,b;x) \; \H(c,d;x) &=&  \H(a,b,c,d;x) + \H(a,c,b,d;x) + \H(a,c,d,b;x) 
                              \nonumber\\
                         &+& \H(c,a,b,d;x)  + \H(c,a,d,b;x) + \H(c,d,a,b;x) 
                             \ , 
\label{eq:alg2} 
\end{eqnarray} 
where $a,b,c,d$ are indices taking any of the values $(1,0,-1)$. 
The formulae can be easily verified, one at a time, 
by observing that they are true at some specific point (such as $x=0$, 
where all the HPLs  vanish except in the otherwise trivial case in which 
all the indices are equal to 0), then taking the $x$-derivatives of the 
two sides according to Eq.\ (\ref{eq:derive}) and checking that they are equal 
(using when needed the previously established lower-weight formulae). 

Another class of identities is obtained by integrating (\ref{eq:defh0}) 
by parts. These integration-by-parts (IBP) identities read:
\begin{eqnarray}
\H(m_1,\ldots,m_q;x) &=&  \H(m_1;x)\H(m_2,\ldots,m_q;x)
                        -\H(m_2,m_1;x)\H(m_3,\ldots,m_q;x) \nonumber \\
&& + \ldots + (-1)^{q+1} \H(m_q,\ldots,m_1;x)\;.
\label{eq:ibp}
\end{eqnarray} 
These identities are not fully linearly 
independent of the product identities.


A numerical implementation of the HPLs up to weight $w=4$ is 
available~\cite{1dhpl}.

\subsection{Two-dimensional harmonic polylogarithms}
The 2dHPLs family 
is obtained by the repeated integration, in the variable $y$, 
of rational factors chosen, in any order, from the set $1/y$, $1/(y-1)$, 
$1/(y+z-1)$, $1/(y+z)$, where $z$ is another independent variable (hence 
the `two-dimensional' in the name). 
In full generality, let us define the rational factor 
as 
\begin{equation} 
  \g(a;y) = \frac{1}{y-a} \ , 
\label{eq:gya} 
\end{equation} 
where $a$ is the {\it index}, which can depend on $z$, $a=a(z)$; 
the rational factors that we consider for the 2dHPLs then are  
\begin{eqnarray} 
  \g(0;y) &=& \frac{1}{y} \ ,       \nonumber\\ 
  \g(1;y) &=& \frac{1}{y-1} \ ,     \nonumber\\ 
  \g(1-z;y) &=& \frac{1}{y+z-1} \ , \nonumber\\ 
  \g(-z;y) &=& \frac{1}{y+z} \ . 
\label{eq:gyalist} 
\end{eqnarray} 
With the above definitions, the index takes one of the values 
$0, 1, (1-z) $ and $(-z)$. \par 
Correspondingly, the 2dHPLs at weight $w=1$ (i.e.~depending, besides 
the variable $y$, on a single further argument, or {\it index}) are 
defined to be 
\begin{eqnarray} 
 \G(0;y) &=& \ln\, y  \ ,                                \nonumber\\ 
 \G(1;y) &=& \ln\, (1-y) \ ,                            \nonumber\\ 
 \G(1-z;y) &=& \ln\left( 1 - \frac{y}{1-z} \right) \ , \nonumber\\ 
 \G(-z;y) &=& \ln\left( 1 + \frac{y}{z} \right) \ . 
\label{eq:w1list} 
\end{eqnarray} 
The 2dHPLs of weight $w$ larger than 1 depend on a set of $w$ indices, which 
can be grouped into a 
$w$-dimensional {\it vector} of indices $\vec{a}$. By writing the vector 
as $ \vec{a} = (a, \vec b)$, where $a$ is the leftmost component of 
$ \vec{a} $ and $\vec b $ stands for the vector of the remaining $(w-1)$ 
components, the 2dHPLs are then defined as follows: if all the $w$ 
components of $\vec a$ take the value 0, $\vec a$ is written as $\vec 0_w$ and 
\begin{equation} 
  \G(\vec{0}_w;y) = \frac{1}{w!} \ln^w{y} \ , 
\label{eq:defg0} 
\end{equation} 
while, if $\vec{a} \neq \vec{0}_w$, 
\begin{equation} 
  \G(\vec{a};y) = \int_0^y \d y' \ \g(a;y') \ \G(\vec{b};y') \ . 
\label{eq:defx0} 
\end{equation} 
In any case the derivatives can be written in the compact form 
\begin{equation} 
\frac{\d }{\d y} \G(\vec{a};y) = \g(a;y) \G(\vec{b};y) \ , 
\label{eq:deriveG} 
\end{equation} 
where, again, $a$ is the leftmost component of $ \vec a $ and 
$ \vec b $ stands for the remaining $(w-1)$ components. 

It should be observed that the notation for the 2dHPLs employed here is the
notation of~\cite{2dhpl}, which is different 
from the original definition proposed in~\cite{doublebox}. Detailed
conversion rules between the different notations, as well as relations to 
similar functions in the mathematical literature (hyperlogarithms and 
multiple polylogarithms) can be found in the appendix of~\cite{2dhpl}. 

Algebra and reduction equations of the 2dHPLs are the same as for the 
ordinary HPLs.
The product of two 2dHPLs  of a same argument $y$ 
and weights $p, q$ can be expressed as a combination of 2dHPLs  of that 
argument and weight $r=p+q$, according to the product identity 
\begin{eqnarray} 
 \G(\vec{p};x)\G(\vec{q};x) & = & 
  \sum_{\vec{r} = \vec{p}\uplus \vec{q}} \G(\vec{r};x) \; , 
\label{eq:galgebra} \end{eqnarray} 
where $\vec p, \vec q$ stand for the $p$ and $q$ components of the indices 
of the two 2dHPLs, while $\vec{p}\uplus \vec{q}$ represents all possible 
mergers of $\vec{p}$ and $\vec{q}$ into the vector $\vec{r}$ with $r$ 
components, in which the relative orders of the elements of $\vec{p}$ 
and $\vec{q}$ are preserved. The explicit product identities up to 
weight $w=4$ are identical to those for the HPLs 
(\ref{eq:alg0})--(\ref{eq:alg2}), with all $\H$ replaced by $\G$. 

The integration-by-parts identities read:
\begin{eqnarray}
\G(m_1,\ldots,m_q;x) &=&  \G(m_1;x)\G(m_2,\ldots,m_q;x)
                        -\G(m_2,m_1;x)\G(m_3,\ldots,m_q;x) \nonumber \\
&& + \ldots + (-1)^{q+1} \G(m_q,\ldots,m_1;x)\;.
\end{eqnarray} 

A numerical implementation of the 2dHPLs up to weight $w=4$ is 
available~\cite{2dhpl}. 

\end{appendix}

\end{document}